\pdfoutput=1

\documentclass[twoside,twocolumn,9pt]{article}
\usepackage{extsizes}
\usepackage[super,sort&compress,comma]{natbib} 
\usepackage[version=3]{mhchem}
\usepackage[left=1.5cm, right=1.5cm, top=1.785cm, bottom=2.0cm]{geometry}
\usepackage{balance}
\usepackage{times,mathptmx}
\usepackage{sectsty}
\usepackage{graphicx} 
\usepackage{lastpage}
\usepackage[format=plain,justification=justified,singlelinecheck=false,font={stretch=1.125,small,sf},labelfont=bf,labelsep=space]{caption}
\usepackage{float}
\usepackage{fancyhdr}
\usepackage{fnpos}
\usepackage[english]{babel}
\addto{\captionsenglish}{%
  
}
\usepackage{array}
\usepackage{droidsans}
\usepackage{charter}
\usepackage[T1]{fontenc}
\usepackage[usenames,dvipsnames]{xcolor}
\usepackage{setspace}
\usepackage[compact]{titlesec}
\usepackage{hyperref}
\usepackage{epstopdf}
\definecolor{cream}{RGB}{222,217,201}

\newcommand{\vvec}[1]{\mathbf{#1}}
\newcommand{\vhat}[1]{\widehat{\mathbf{#1}}}
\newcommand{\idx}[1]{_{\mathrm{#1}}}
\newcommand{\upx}[1]{^{\mathrm{#1}}}
\newcommand{\ns}{\textrm{ns}}
\newcommand{\nm}{\textrm{nm}}
\newcommand{\J}{\textrm{J}}
\newcommand{\fs}{\,\mathrm{fs}}
\newcommand{\gcm}{\,\mathrm{g}/\mathrm{cm}^3}
\newcommand{\K}{\,\mathrm{K}}

\begin{document}

\pagestyle{fancy}
\thispagestyle{plain}
\fancypagestyle{plain}{

\fancyhead[C]{\includegraphics[width=18.5cm]{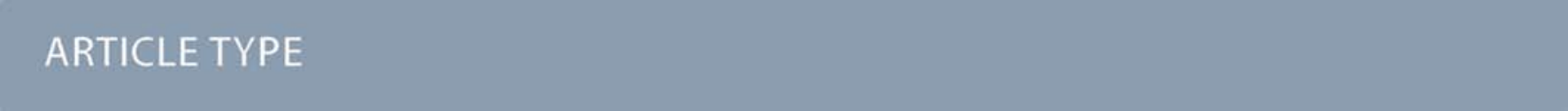}}
\fancyhead[L]{\hspace{0cm}\vspace{1.5cm}\includegraphics[height=30pt]{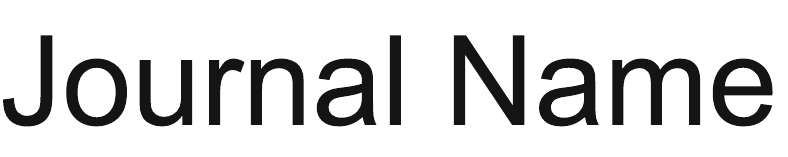}}
\fancyhead[R]{\hspace{0cm}\vspace{1.7cm}\includegraphics[height=55pt]{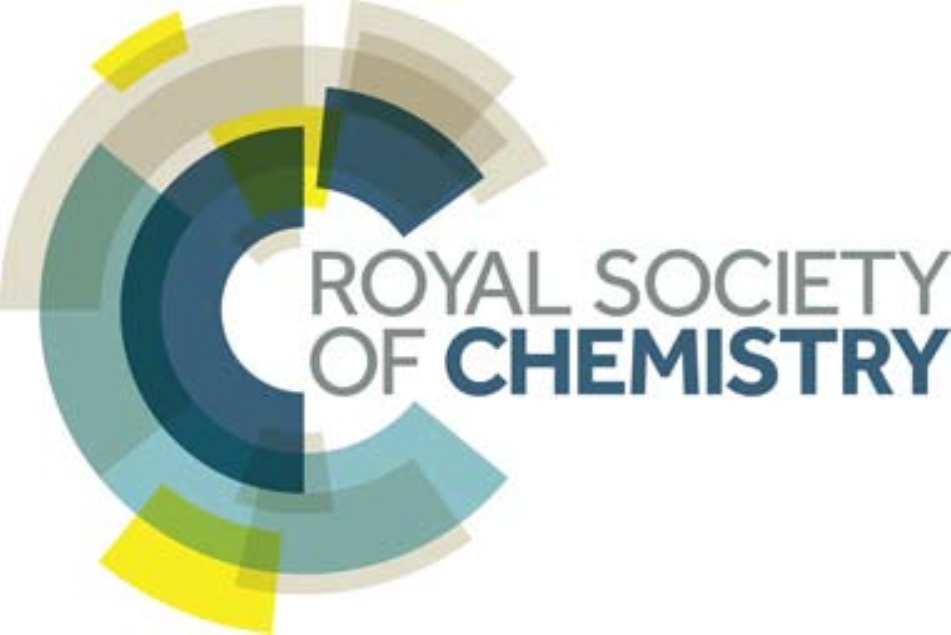}}
\renewcommand{\headrulewidth}{0pt}
}

\makeFNbottom
\makeatletter
\renewcommand\LARGE{\@setfontsize\LARGE{15pt}{17}}
\renewcommand\Large{\@setfontsize\Large{12pt}{14}}
\renewcommand\large{\@setfontsize\large{10pt}{12}}
\renewcommand\footnotesize{\@setfontsize\footnotesize{7pt}{10}}
\makeatother

\renewcommand{\thefootnote}{\fnsymbol{footnote}}
\renewcommand\footnoterule{\vspace*{1pt}%
\color{cream}\hrule width 3.5in height 0.4pt \color{black}\vspace*{5pt}} 
\setcounter{secnumdepth}{5}

\makeatletter 
\renewcommand\@biblabel[1]{#1}            
\renewcommand\@makefntext[1]%
{\noindent\makebox[0pt][r]{\@thefnmark\,}#1}
\makeatother 
\renewcommand{\figurename}{\small{Fig.}~}
\sectionfont{\sffamily\Large}
\subsectionfont{\normalsize}
\subsubsectionfont{\bf}
\setstretch{1.125} 
\setlength{\skip\footins}{0.8cm}
\setlength{\footnotesep}{0.25cm}
\setlength{\jot}{10pt}
\titlespacing*{\section}{0pt}{4pt}{4pt}
\titlespacing*{\subsection}{0pt}{15pt}{1pt}

\fancyfoot{}
\fancyfoot[LO,RE]{\vspace{-7.1pt}\includegraphics[height=9pt]{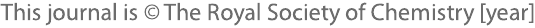}}
\fancyfoot[CO]{\vspace{-7.1pt}\hspace{13.2cm}\includegraphics{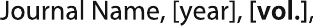}}
\fancyfoot[CE]{\vspace{-7.2pt}\hspace{-14.2cm}\includegraphics{head_foot/RF}}
\fancyfoot[RO]{\footnotesize{\sffamily{1--\pageref{LastPage} ~\textbar  \hspace{2pt}\thepage}}}
\fancyfoot[LE]{\footnotesize{\sffamily{\thepage~\textbar\hspace{3.45cm} 1--\pageref{LastPage}}}}
\fancyhead{}
\renewcommand{\headrulewidth}{0pt} 
\renewcommand{\footrulewidth}{0pt}
\setlength{\arrayrulewidth}{1pt}
\setlength{\columnsep}{6.5mm}
\setlength\bibsep{1pt}

\makeatletter 
\newlength{\figrulesep} 
\setlength{\figrulesep}{0.5\textfloatsep} 

\newcommand{\topfigrule}{\vspace*{-1pt}%
\noindent{\color{cream}\rule[-\figrulesep]{\columnwidth}{1.5pt}} }

\newcommand{\botfigrule}{\vspace*{-2pt}%
\noindent{\color{cream}\rule[\figrulesep]{\columnwidth}{1.5pt}} }

\newcommand{\dblfigrule}{\vspace*{-1pt}%
\noindent{\color{cream}\rule[-\figrulesep]{\textwidth}{1.5pt}} }

\makeatother

\twocolumn[
  \begin{@twocolumnfalse}
\vspace{3cm}
\sffamily
\begin{tabular}{m{4.5cm} p{13.5cm} }

\includegraphics{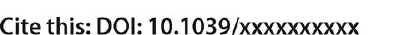} & \noindent\LARGE{\textbf{Gelation of patchy gold nanoparticles
decorated by liquid-crystalline ligands: computer simulation study}} \\
\vspace{0.3cm} & \vspace{0.3cm} \\
 & \noindent\large{Jaroslav M. Ilnytskyi,\textit{$^{a,b}$} Arsen Slyusarchuk,\textit{$^{b}$} and
                   Stefan Soko{\l}owski\textit{$^{c}$}} \\

                   \includegraphics{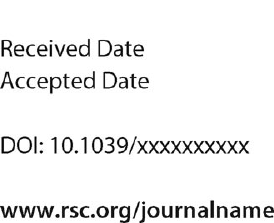} & \noindent\normalsize{
We consider patchy gold nanoparticles decorated by liquid crystalline ligands.
The cases of two, three, four and six symmetrically arranged patches of ligands are discussed,
as well as the cases of their equatorial and uniform arrangement. A solution of decorated
nanoparticles is considered within a flat pore with the solid walls and the interior filled by
a polar solvent. The ligands form physical crosslinks between the nanoparticles due to strong
liquid crystalline interaction, turning the solution into a gel-like structure. Gelation is done
repeatedly starting each time from freshly equilibrated dispersed state of nanoparticles.
The gelation dynamics and the range of network characteristics of gel are examined, depending on
the type of patchy decoration and the solution density. The emphasis is given to the suitability
of a gel for catalytic applications} \\

\end{tabular}
 \end{@twocolumnfalse} \vspace{0.6cm}
  ]

\renewcommand*\rmdefault{bch}\normalfont\upshape
\rmfamily
\section*{}
\vspace{-1cm}


\footnotetext{\textit{$^{a}$~Institute for Condensed Matter Physics of the National Academy of Sciences of Ukraine,
1, Svientsitskii Str., 79011 Lviv, Ukraine. Fax: +38 032 2761158; Tel: +38 032 2761978; E-mail: iln@icmp.lviv.ua}}
\footnotetext{\textit{$^{b}$~National University Lviv Politechnic, 12, Bandera Str., 79000 Lviv, Ukraine.}}
\footnotetext{\textit{$^{c}$~Maria Curie-Sk{\l}odowska University, 33, Gliniana Str., Lublin, Poland.}}






\section{\label{I}Introduction}

Gold nanoparticles (GNP) are important components of a number of different applications.
Examples include the surface plasmon resonance based applications,\cite{Amendola2017}
catalysis,\cite{Hvolbk2007,Thompson2007,Alshammari2016} as well as medical diagnostics
and therapeutics.\cite{Yeh2012,Das2011} Among the conventional methods of synthesis of GNPs
is the reduction of gold(III) derivatives, e.g. using citrate reduction of $HAuCl_4$ in
water, introduced by Turkevich in 1951.\cite{Turkevich1951}
A number of other approaches that exist since then  were reviewed in detail in Sec.3 of
Ref.~\cite{Daniel2004}

The aggregation of GNPs in solution is affected by cationic and oligocationic species and can
be achieved by addition of a specific linker.\cite{Pranami2009,Gerth2017} In particular,
the linear aggregates of GNPs were obtained due to preferential binding of a cetyltrimethylammonium
bromide linker molecules on a certain facet of GNP and GNP-GNP electrostatic interactions.
\cite{Yang2007} The organic linkers containing both thiol and amine groups were found to
strongly promote the aggregation of citrate-stabilized GNPs into single, chain-like, and
globular structures. \cite{Chegel2012} Incorporation of photosensitive groups into a linker,
e.g. the azobenzene, allows to achieve a photo-controllable aggregation of GNPs, as
shown in Refs.~\cite{Klajn2007,Kawai2008,Wei2010,Lysyakova2015}

One of the most technologically important applications of GNPs is the heterogeneous catalysis.
\cite{Hvolbk2007,Thompson2007,Alshammari2016} Gold has been always considered chemically inert
until it was discovered that the GNP of the $<5\nm$ sizes can be very effective catalysts.
\cite{Haruta1987} As remarked in Ref.~\cite{Hvolbk2007}, in some cases catalysts based on GNPs,
allow for a significantly  lower reaction temperature than that used in typical processes.
The heterogeneous catalysis is based on adsorption of reacting molecules on the
catalytically active solid surface. The chemical bonds between reagents are then formed
on the surface with consequent release of the final product of reaction into the host phase.

The effectiveness of catalysis depends on a number of factors such as accessible surface area
of GNP, diffusivity of reacting species and of the final product of reaction, etc.
\cite{Hvolbk2007,Thompson2007,Alshammari2016} These factors can be greatly enhanced 
by using nanoporous metals characterized by tunable porosity and high structural stability.
\cite{Fujita2017,Christiansen2018} Another class of materials, GNP-containing gels, possess
similar properties and are cheaper to fabricate.
\cite{Corain2010,Ramtenki2012,Che2015,Wu2015,Wen2016,Zhang2016,Wei2018,Pia2018}

One of the ways to form a GNP-containing gel is to crosslink GNPs decorated by suitable ligands.
\cite{Klajn2007,Kawai2008,Pranami2009,Wei2010,Lysyakova2015} The current state in synthesis
of GNPs allows their precise decoration at particular surface points or areas leading to the
overally discotic shape,\cite{Kumar2005} as well as poles plus equator decorated GNPs. \cite{Dias2013}
Other patching patterns are possible allowing tunability of such properties of the resulting
gel as pores distribution and elasticity relevant to catalytic applications.\cite{Kazem2015,Pia2018}

In this study we consider a range of model GNPs that are decorated by ligands using various patching
patterns, termed hereafter as patchy gold nanoparticles (PGNP). Each ligand consists of a short chain
terminated by a liquid crystalline (LC) group. In solution, the LC groups attract each other providing
the means of physical crosslinks and resulting in gel formation. We focus on the dependence of the
gelation dynamics and the properties of the gel on the pattern type and on the solution concentration.
The study to some extend combines and generalizes some of the previous results on bulk morphologies
of decorated GNPs\cite{Ilnytskyi2010,Ilnytskyi2013,Slyusarchuk2014} and on their gelation
\cite{Ilnytskyi2016a,Ilnytskyi2017} with the studies of ``hairy'' particles near the walls.
\cite{Baran2017a,Baran2017b}

The outline of the study is as follows. In Sec.~\ref{II} we describe in detail the model PGNP
with a variety of decoration patterns and discuss the interaction potentials. Sec.~\ref{III}
contains the results for gelation dynamics at fixed solution concentration. In Sec.~\ref{IV}
we discuss the properties of a gel at various concentration of a solution, whereas the conclusions
are provided in Sec.~\ref{V}.

\section{\label{II}Models for patchy decorated gold nanoparticles}

%
\begin{figure}[!th]
\begin{center}
\includegraphics[clip,width=7cm]{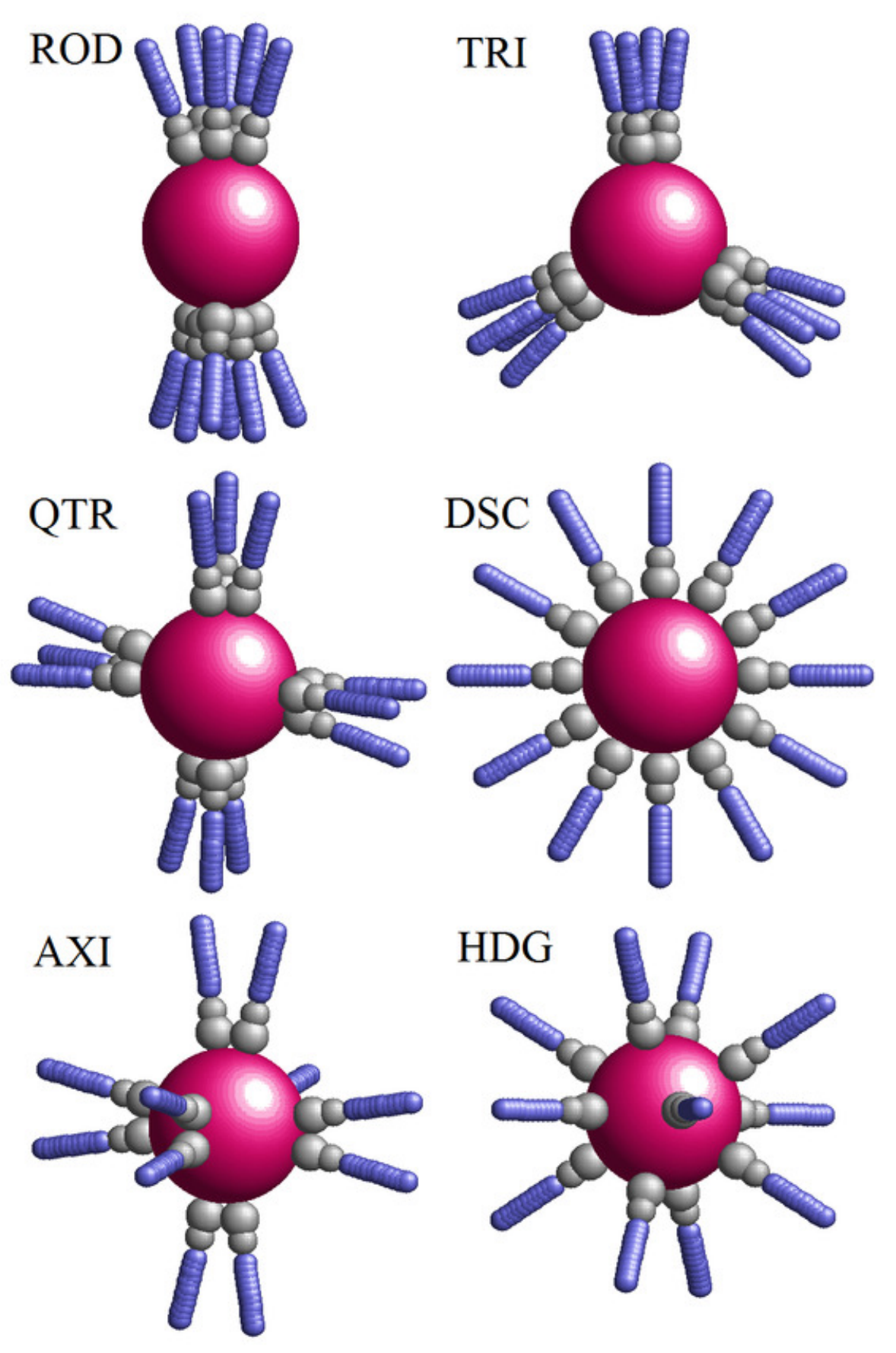}
\caption{\label{model}Classification of PGNP used in this study.
The GNP is shown as a large pink sphere, each ligand contains a spacer of two
spherical beads (shown in gray) and the azobenzene chromophore (shown as a blue
spherocylinder).}
\end{center}
\end{figure}

In this study we use the coarse-grained type of modelling for the PGNP, where the GNP is represented as a
large spherical bead, the polymer part of a ligand -- as two spherical beads of smaller
size and the LC groups -- as a spherocylinder bead. Therefore, all beads represent a collection
of atoms and interact via effective potentials. More discussion can be found elsewhere.
\cite{Ilnytskyi2010,Ilnytskyi2013,Slyusarchuk2014}
To cover various types of decoration of GNP by LC ligands, we developed 
a set of
new coarse-grained models classified in Fig.~\ref{model}. These 
extend the model considered
in Refs.~\cite{Ilnytskyi2010,Ilnytskyi2013,Ilnytskyi2016a,Ilnytskyi2016b,Ilnytskyi2017}
by applying additional constraints to the movement of ligands with respect to the GNP,
by employing the approach from Refs.~\cite{Baran2017a,Baran2017b}

Let us first describe the models shown in Fig.~\ref{model} qualitatively, whereas the potential
energy form will be given below. All models have the same number of ligands equal to $N\idx{l}=12$.
The model ROD is of $1D$ patching symmetry with two patches located at the polar regions
of GNP, each formed of $6$ ligands.
The first bead of each ligand is grafted to the center of a GNP by a harmonic bond.
To keep the ligands that belong to the same patch together, we introduce bonds connecting
their respective grafted beads. Then, to ensure polar arrangement of the two patches, we use
harmonic angle [1]-[O]-[2], where [1] is one of the grafted beads of first patch, [O] is the
center of the central sphere, and [2] is one of the grafted beads of the second patch. The reference 
angle is set to $\theta'_0=\pi$. As the result, the grafted beads of ligands are able to slide on
the surface of a GNP, but in a form of a relatively rigid scaffold. Due to spherical symmetry
of a central sphere, such sliding motion is energetically equivalent to the rotation of the
experimental decorated GNP, in which case grafting occurs as fixed points of the core GNP.

The TRI and QTR models are constructed in a similar way. The former has three patches of
$4$ ligands each and three interpatch angles [1]-[0]-[2], [2]-[0]-[3] and [3]-[0]-[1] with
$\theta'_0=2\pi/3$. The latter has four patches of $3$ ligands each with four angles [1]-[0]-[2],
[2]-[0]-[3], [3]-[0]-[4] and [4]-[0]-[1] and $\theta'_0=\pi/2$ plus two additional angles
[1]-[0]-[3] and [2]-[0]-[4] with $\theta'_0=\pi$. Both models have $2D$ patching
symmetry, and their natural continuation is the EQU model with twelve single ligand patches arranged
equidistantly on the equator of the GNP. We found that the ligands keeps better equatorial circle
throughout the simulation when the following twelve angles are used: [1]-[2]-[3], [2]-[3]-[4],
$\cdots$, [12]-[1]-[2] with $\theta'_0=5\pi/6$. This patching pattern is reminiscent of discotic-decorated
GNPs discussed in Ref.~\cite{Kumar2005}

Finally, the models AXI and HDG, have $3D$ patching symmetry. In the former, six patches of $2$
ligands each are arranged along each of three axes and the interpatch angles are introduced
similarly to the case of the QTR model. In the HDG model, there are twelve single ligand patches
arranged uniformly on the surface of the GNP, namely, on the vertices of a regular icosahedron
resulting in a hedgeholg-like appearance. The angles [m]-[0]-[n] are introduced for each pair
(m,n) of vertices that form an edge, totaling in $30$.

The expression for the bonded interactions written for a single molecule has the following form:
\begin{eqnarray}\label{Vb}
 V_{b}&=&\sum_{i=1}^{N_b}k_b(l_i-l_0)^2 +
       \sum_{i=1}^{N_a}k_a(\theta_{i}-\theta_0)^2\\
       &+& \sum_{i=1}^{N'_a}k'_a(\theta_{i}-\theta'_0)^2 +
       \sum_{i=1}^{N_z}k_z(\zeta_{i}-\zeta_0)^2,
\end{eqnarray}
where $N_b$, $N_a$, $N''a$ and $N_z$ are the total numbers of bonds, intraligand angles, interpatch angles
and ligands in a single molecule. The last term ensures that the LC group is connected properly to the
ligand spacer.\cite{Wilson2003} The following force field constants are used:
$k_b=50\cdot 10^{-20}\J/(0.1\nm)^2$;
$l_0=1.49\nm$ (GNP-to-first spacer bead), $l_0=0.36\nm$ (first-to-second spacer bead)
and $l_0=0.859\nm$ (second spacer bead-to-center of the LC group); $k_a=20\cdot 10^{-20}\J/\mathrm{rad}^2$ and
$k'_a=50\cdot 10^{-20}\J/\mathrm{rad}^2$. Bending reference angles are $\theta'_0=\pi$ for the intraligand bending
potential, whereas the interpatch ones, $\theta'_0$, depend on the decoration option and are specified above.

\begin{figure}
\begin{center}
(a)\vspace{-1.5em}\\
\includegraphics[clip,angle=270,width=6cm]{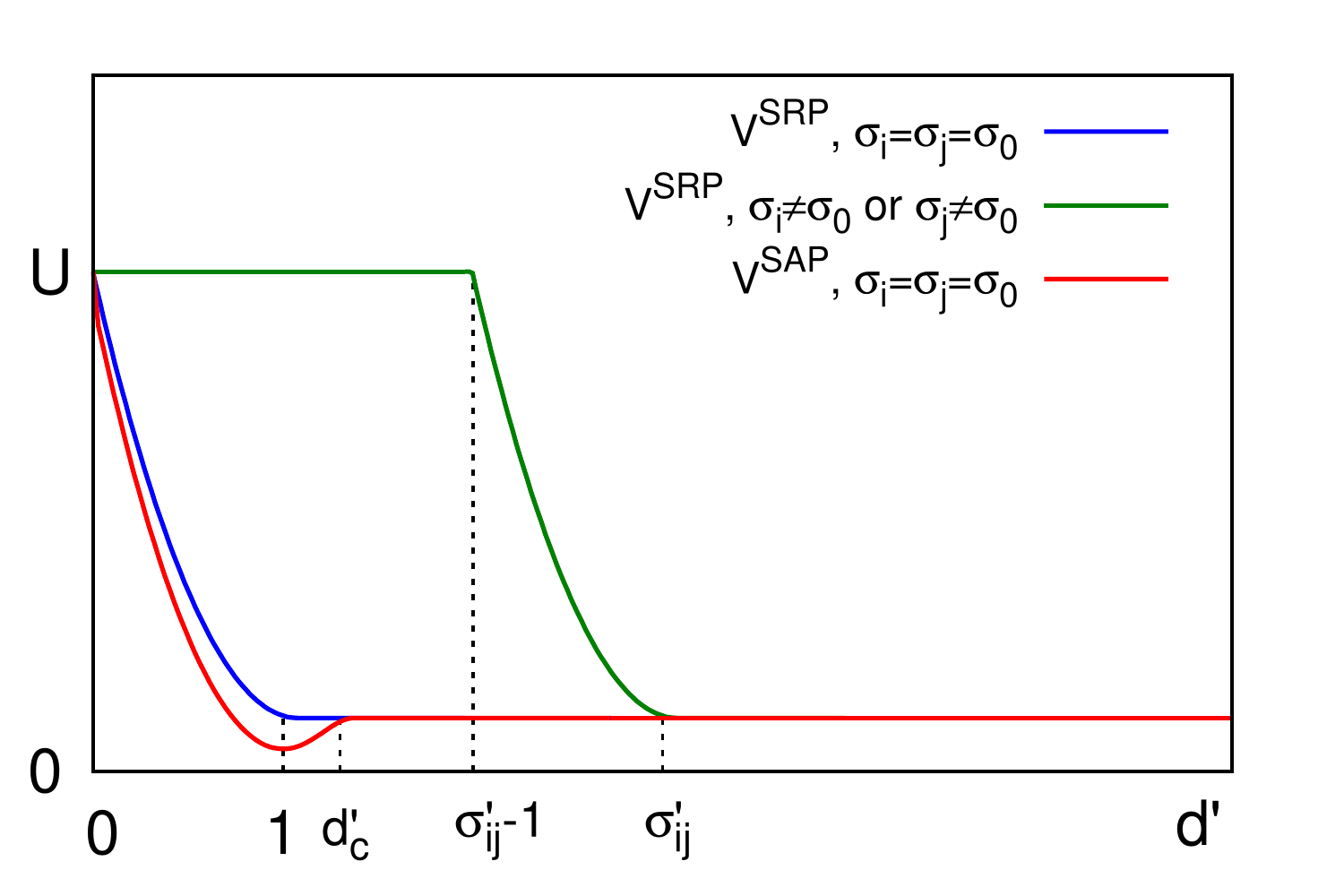}\\
(b)\vspace{-1.5em}\\
\includegraphics[clip,angle=270,width=6cm]{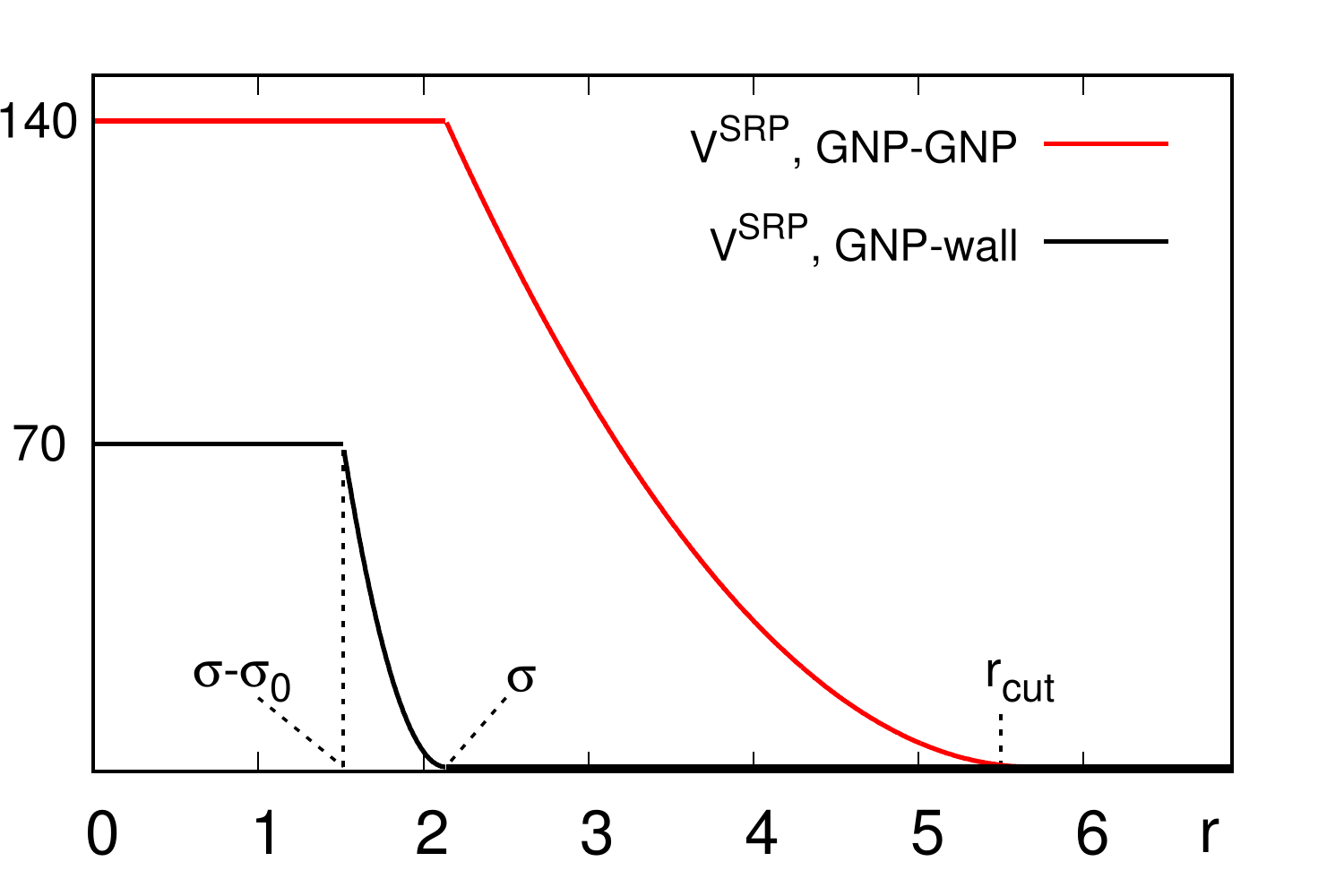}
\caption{\label{potent}(a) SRP for the cases of equal and unequal beads dimensions, $\sigma_1$ and
$\sigma_2$, and SAP for the case $\sigma_1=\sigma_2$, $d'$ is reduced separation between two cores, $U$ -- the
energy scale. (b) Comparison for the SRP for the GNP-GNP interaction with that for the GNP-wall one using chosen
energy scales of $U=140\cdot 10^{-20}\J$ and $U=70\cdot 10^{-20}\J$, respectively.}
\end{center}
\end{figure}
All types of nonbonded potentials used in this study can be obtained from just two analytic forms:
soft repulsive potential (SRP) and soft attractive potential (SAP). They are defined with respect
to the convex cores of the beads, according to Kihara.\cite{Kihara1963} Namely, the convex core 
for a spherocylinder is a line
that connects the centers of its two spherical caps, whereas for the sphere it reduces
itself to the center of a sphere.
Let $\vvec{r}_{ij}$ be the vector that connects the centers of $i$th and $j$th beads and $\vhat{e}_i$
and $\vhat{e}_j$ are their orientations [they are omitted for spherical bead(s)]. Then, the set
$\vvec{q}_{ij}=\{\vhat{e}_i,\vhat{e}_j,\vvec{r}_{ij}\}$ characterizes the mutual spatial position of
the $i$th and $j$th beads and the shortest distance between their cores is denoted as $d(\vvec{q}_{ij})$.
The SRP can be written in a scaled  form as
\begin{equation}\label{SRP}
\!V\upx{SRP}\!=\!\left\{\!
\begin{array}{ll}
U, &\!\!d'(\vvec{q}_{ij})<\sigma'_{ij}-1\\
U\left[\!1\!-\![d'(\vvec{q}_{ij})\!-\!\sigma'_{ij}\!+\!1]\right]^2\!\!, &\!\!\sigma'_{ij}\!-\!1\leq\!d'(\vvec{q}_{ij})\!\leq\!\sigma'_{ij}\vspace{2mm}\\
0, &\!\!d'(\vvec{q}_{ij}) > \sigma'_{ij}
\end{array}
\right.,
\end{equation}
where $U$ is the energy scale, $d'(\vvec{q}_{ij}=d(\vvec{q}_{ij}\sigma_0$ is reduced shortest distance between the cores,
$\sigma'_{ij}=(1/2)(\sigma_i+\sigma_j)/\sigma_0$ is the mean value of the characteristic dimensions of two beads, and
$\sigma_0$ is the length scale. The diameters of the spherical beads are: $\sigma=2.137\nm$ (the GNP), $0.623\nm$
(spacer bead grafted to GNP), $0.459\nm$ (middle spherical bead of a spacer and solvent beads). The LC groups are
represented as spherocylinders with the breadth of $D=0.374\nm$ and the elongation of $L/D=3$. These dimensions
are estimated in some earlier works on coarse-graining of similar macromolecular systems.\cite{Hughes2005}
The energy scale is set as $U=70\cdot 10^{-20}\J$
For all the pairs of spherical particles, except the GNP-GNP pair, we set the energy scale at $U=70\cdot 10^{-20}\J$
and the length scale at $\sigma_0=0.459\nm$ (the smallest spherical particles in a model).
Therefore, in general case, one obtains a shifted form of the interaction potential for such pairs of beads, 
as shown in Fig.~\ref{potent} (a)
and marked there as $V\upx{SRP},\sigma_i\neq\sigma_0 \mathrm{\ or \ } \sigma_j\neq\sigma_0$.
We opted for this form to avoid softening
of the repulsive potential towards the edge of a GNP, which is the case for 
the scaled only form of the interaction potential.
The shift disappears in a special case of $\sigma_i=\sigma_j=\sigma_0$ shown in the same figure.
If one of the beads, e.g. $i$th one, is of the LC type then we set $\sigma_i=D$, while keeping the same $\sigma_0=0.459\nm$
as for the pair of spherical particles. If both beads are of the LC type then $\sigma_i=\sigma_j=\sigma_0=D$. The form
of the potential between two LC beads, written as a function of dimensionless $d'(\vvec{q}_{ij})$, is the same as
for the spherical particles at $\sigma_i=\sigma_j=\sigma_0$ mentioned above.

The general expression for the attractive potential can be written as\cite{Lintuvuori2008,Ilnytskyi2014,Ilnytskyi2015}
\begin{equation}\label{SAP}
V\upx{SAP}=\left\{
\begin{array}{ll}
\!\!\!\!\!\!U, &\!\!\!\!\!\!d'(\vvec{q}_{ij})<\sigma'_{ij}-1\\
\!\!\!\!\!\!U\big\{[1-d'(\vvec{q}_{ij})-\sigma'_{ij}+1]^2-\epsilon'(\vvec{q}_{ij})\big\},&\!\!\!\!\!\! d'(\vvec{q}_{ij})\in[\sigma'_{ij}-1,\sigma'_{ij}]\vspace{2mm}\\
\!\!\!\!\!\!U\big\{[1-d'(\vvec{q}_{ij})-\sigma'_{ij}+1]^2-\epsilon'(\vvec{q}_{ij})\big. &\vspace{2mm}\\
  \big.\hspace{0.5em}-\frac{1}{4\epsilon'(\vvec{q}_{ij})}[1-d'(\vvec{q}_{ij})-\sigma'_{ij}+1]^4\big\},&\!\!\!\!\!\! d'(\vvec{q}_{ij})\in[\sigma'_{ij},d'_c]\vspace{2mm}\\
\!\!\!\!\!\!0,&\!\!\!\!\!\! d'(\vvec{q}_{ij})>d'_{c}
\end{array}
\right.
\end{equation}
\begin{equation}\label{eps}
\epsilon'(\vvec{q}_{ij})=
\Bigg\{4\Big[
U'_a
-5\epsilon'_1 P_2(\vhat{e}_i\cdot\vhat{e}_j)
-5\epsilon'_2\Big(P_2(\vhat{r}_{ij}\cdot\vhat{e}_i)+P_2(\vhat{r}_{ij}\cdot\vhat{e}_j)\Big)
\Big]\Bigg\}^{-1},
\end{equation} 
where the dimensionless configuration dependent well depth $\epsilon'(\vvec{q}_{ij})$ is obtained from the requirement
that both the potential and its first derivative vanish at $d'_{c}$.\cite{Lintuvuori2008}
Here $\vhat{r}_{ij}=\vvec{r}_{ij}/r_{ij}$ is the unit vector along the vector $\vvec{r}_{ij}$,
$U'_a$, $\epsilon'_1$ and $\epsilon'_2$ are dimensionless parameters that define the shape of the potential,
$P_2(x)=(3x^2-1)/2$ is the second Legendre polynomial. The shape of this potential is illustrated
for the case of the same type of interacting beads in Fig.~\ref{potent} (a) marked as
$V\upx{SAP},\sigma_i=\sigma_j=\sigma_0$. For the case of different bead types, it has a shifted form
with the well (not shown).

The case of GNP-GNP pairwise interactions is a special one. Typical stabilization agent of GNP comprises
a double electric layer with negatively charged exterior.\cite{Lysyakova2015} In this case the stabilized GNP are
subject to electrostatic repulsion and their aggregation is hampered resulting in the dispersed state.
The most accurate way to take this feature into account would be to introduce the charged particles
into the simulations explicitly. However, to simplify the simulation model, in this study we use a simpler
approach, namely using strong GNP-GNP repulsion of a range $r\idx{cut}=5.5\nm$ with the $\sigma_0$ parameter
equal to $2.137\nm$ (the GNP diameter) and maximum repulsion energy of $U=140\cdot 10^{-20}\J$, indicated as 
as $V\upx{SRP}, \mathrm{GNP-GNP}$ in Fig.~\ref{potent} (b), where it is shown in physical units.
The choice of parameters was
made aiming to achieve relatively long-ranged slowly decaying repulsive interaction. Its range, $r\idx{cut}$,
however, can not exceed a half of the smallest dimension of a simulation box. The separate simulation was
undertaken for the solution of unmodified (bare) GNP particles to confirm that such repulsive interaction
between them prevents their aggregation.

The simulation setup for a pore filled by a solution of GNPs is as follows. The simulation box has the
dimensions of $L_x=13.15\nm$, $L_y=11.39\nm$ and $L_z=20\nm$ along the respective spatial axes.
To mimic a slit-like pore, both bottom, $Z=0$, and top, $Z=L_z$, walls are decorated by a $6\times6$
monolayer of spherical particles of the same type as the core of the patchy GNPs shown in Fig.~\ref{model}.
The wall particles are arranged regularly, as a nearly close-packed hexagonal lattice, 
with the lattice spacing of $2.19\nm$.
The interior of the pore is filled by a required number $N\idx{mol}$ of patchy GNPs of the same type
(one of those shown in Fig.~\ref{model}) at random positions, then the remaining volume is filled by a solvent.
The total solute and solvent density within an accessible volume of a pore is kept constant at about $0.53\gcm$.
Due to a soft nature of both pairwise potentials, Eqs.~(\ref{SRP}) and (\ref{SAP}), the overlaps do not
produce gigantic energies and forces, as would be in the case of atomic simulation with e.g. Lennard-Jones
interaction potentials. Therefore, one can simplify the system preparation by avoiding a two step procedure
of generating a low density system first and then compressing it to required density.

The wall particles are frozen and interact with all the other particle types via the repulsive potential
$V\upx{SRP}$, cf. Eq.~(\ref{SRP}). For the case of the interaction between the GNP and wall particle its plot,
indicated there as $V\upx{SRP}, \mathrm{GNP-wall}$, is shown in bottom frame of Fig.~\ref{potent}.
All simulations are performed in the $NVT$ ensemble at the temperature of $T=480\K$. This choice is based on
a number of previous studies on the self-assembly of similar coarse-grained models.
\cite{Ilnytskyi2010,Ilnytskyi2013,Ilnytskyi2016a,Ilnytskyi2016b} In particular,
the liquid crystallinity in such models disappears at approximately $500-510\K$ and the temperatures 
within the range of $480-490K$ are found to be best suited for the LC-based aggregation\cite{Ilnytskyi2016a}
or spontaneous self-assembly.\cite{Ilnytskyi2016b}
This is exactly the range of temperatures that are high enough to ensure sufficient mobility of PGNPs and,
instantaneously, are not too high to hamper their liquid crystallinity, which is the principal mechanism for
gelation of PGNPs. The temperature control is provided by means of velocity rescaling.
Due to soft nature of all interaction potentials, the time-step of the simulations is chosen
equal to $20\fs$, much longer than in a typical atom-based molecular dynamics simulation.

\section{\label{III}Gelation dynamics and the properties of a gel network at fixed selected concentration of PGNPs}

\begin{figure}
\begin{center}
ROD\hspace{8em}TRI\\
\includegraphics[clip,angle=0,height=4cm]{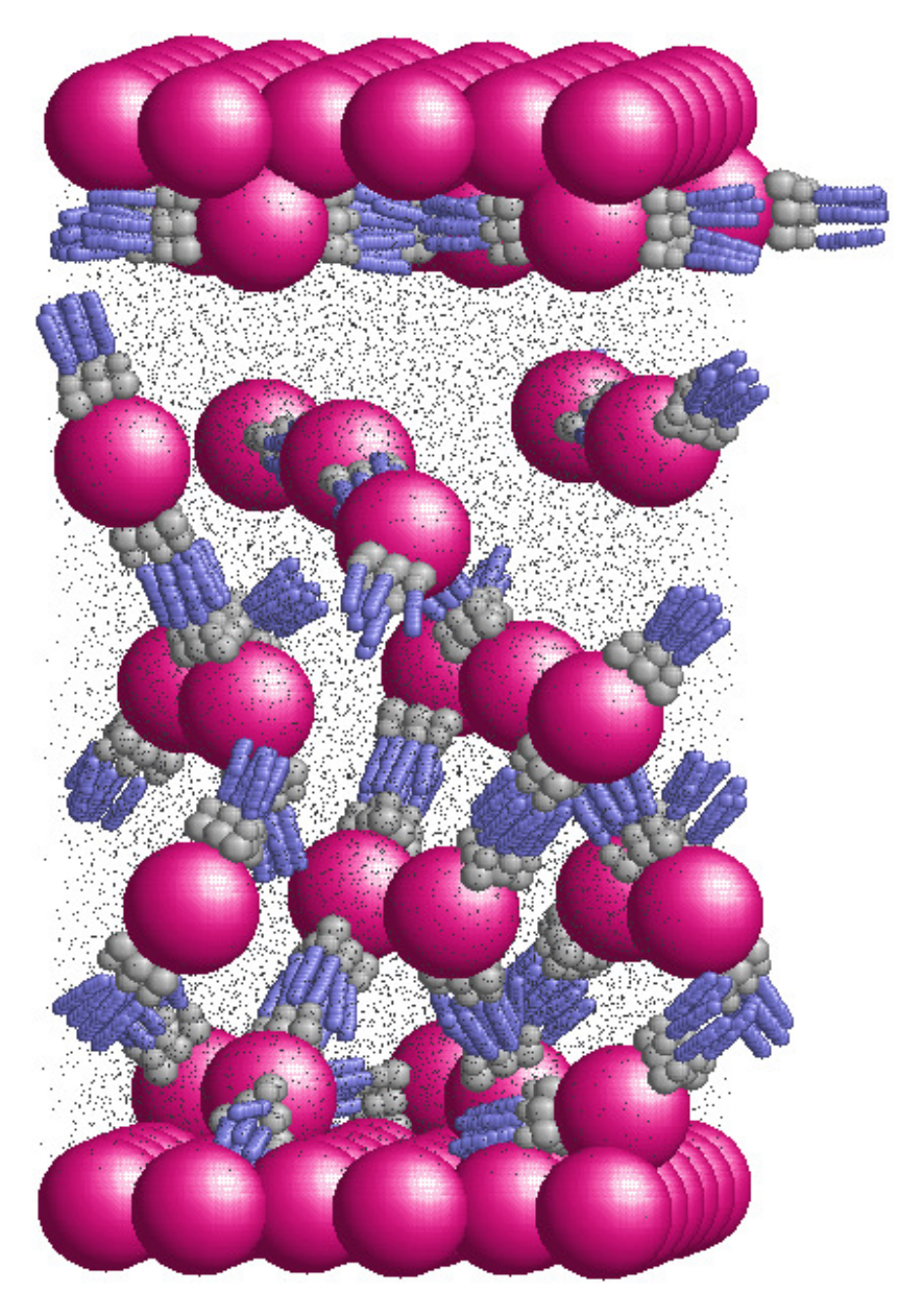}
\includegraphics[clip,angle=0,height=4cm]{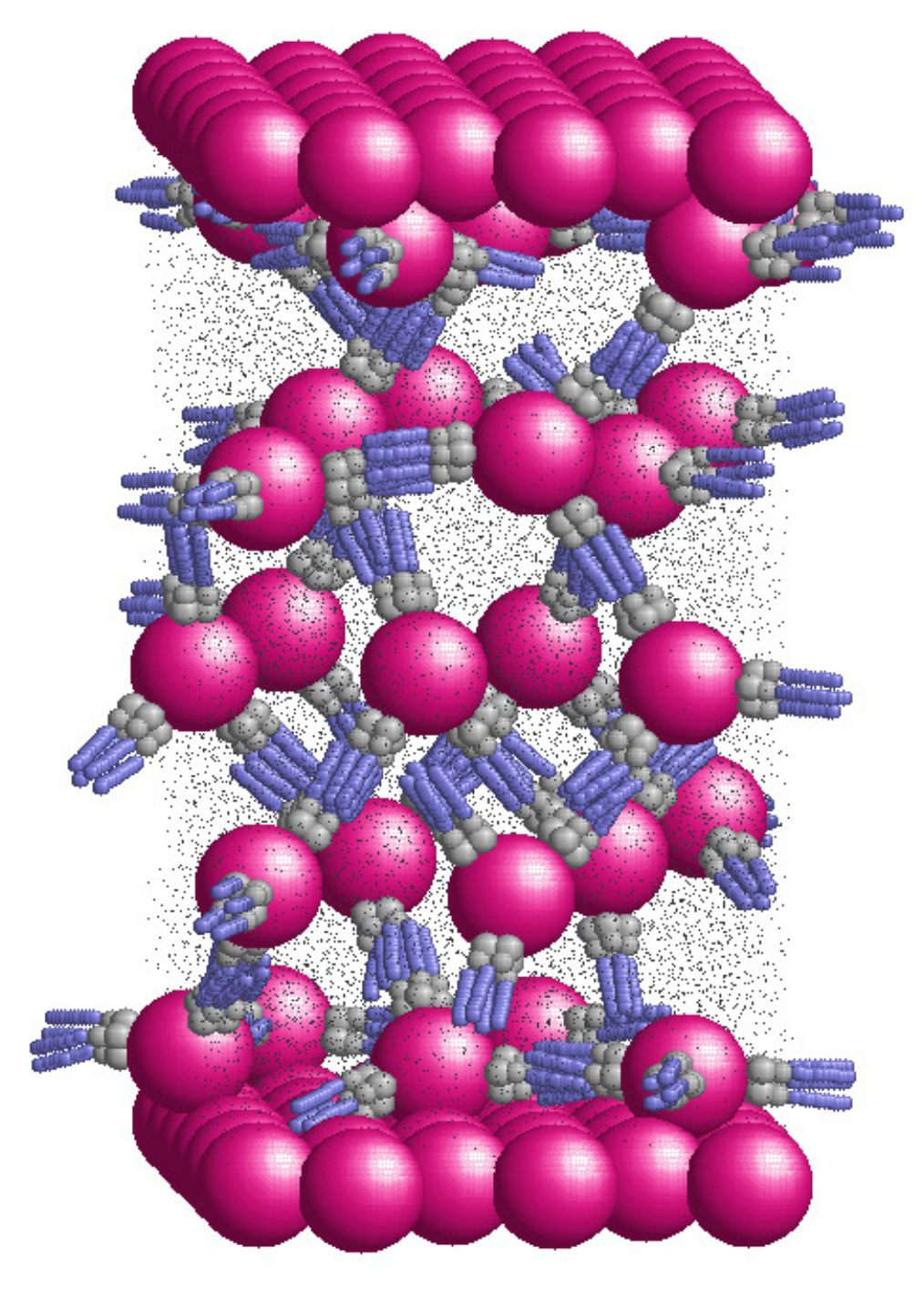}\\
QTR\hspace{8em}DSC\\
\includegraphics[clip,angle=0,height=4cm]{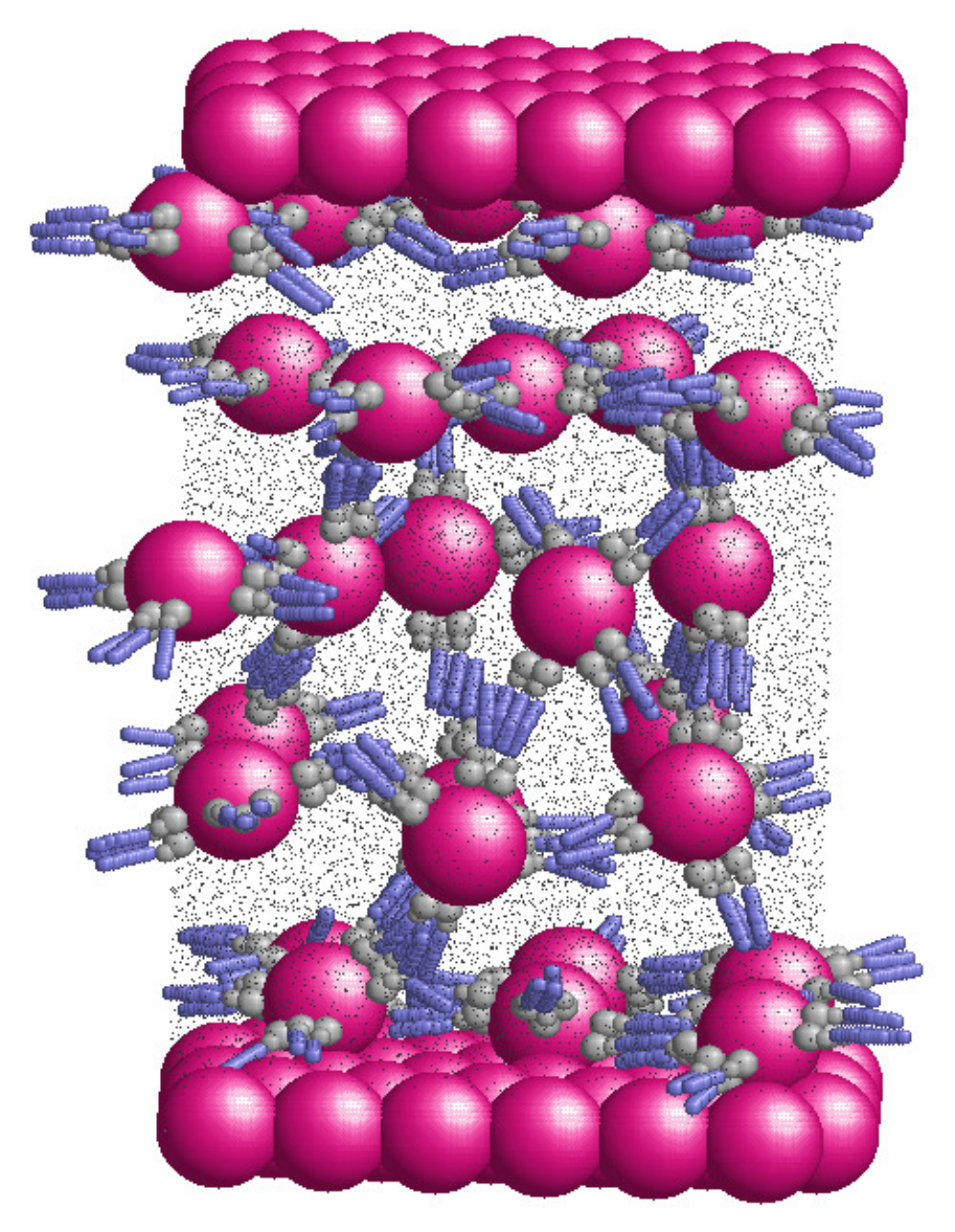}
\includegraphics[clip,angle=0,height=4cm]{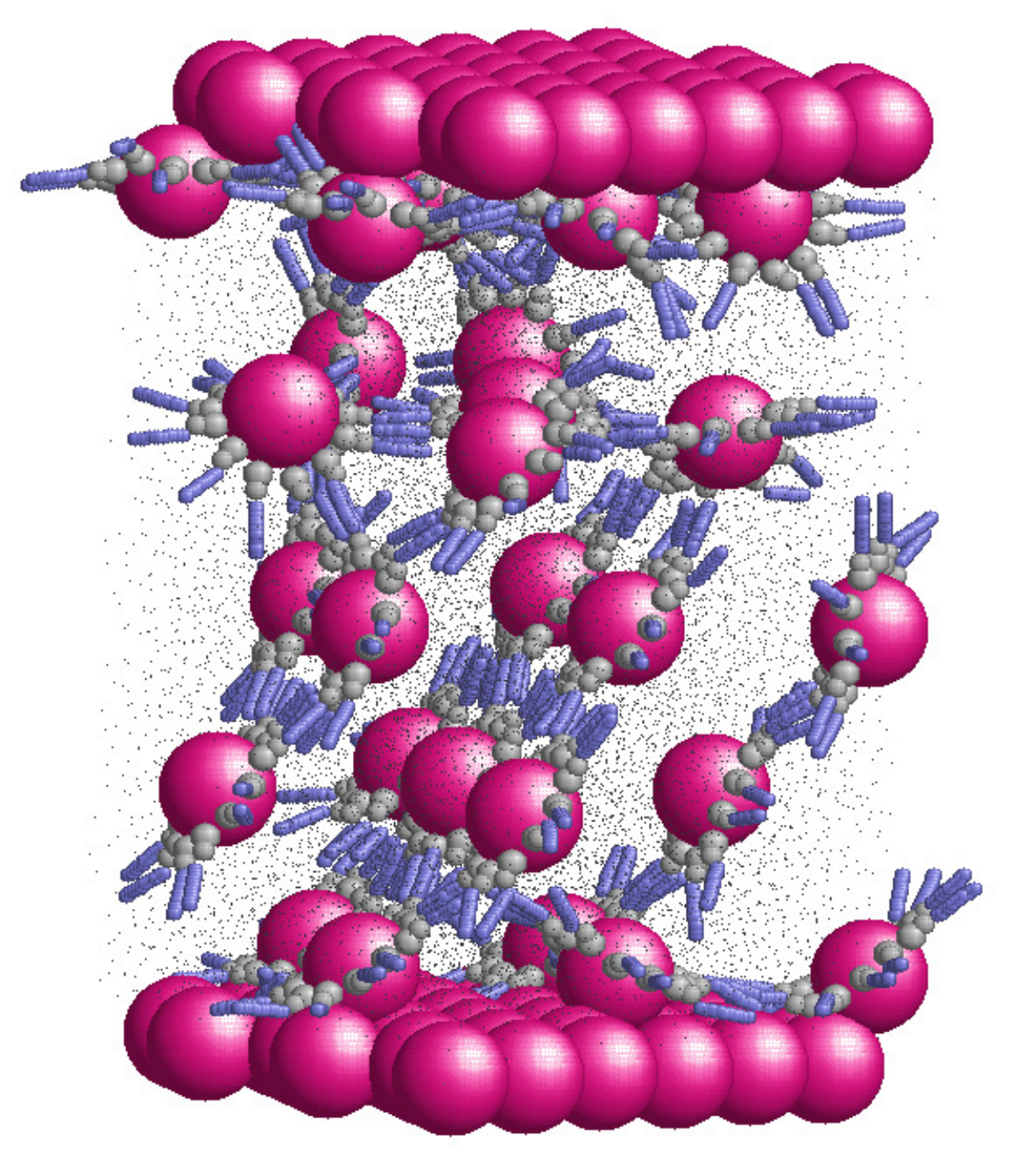}\\
AXI\hspace{8em}HDG\\
\includegraphics[clip,angle=0,height=4cm]{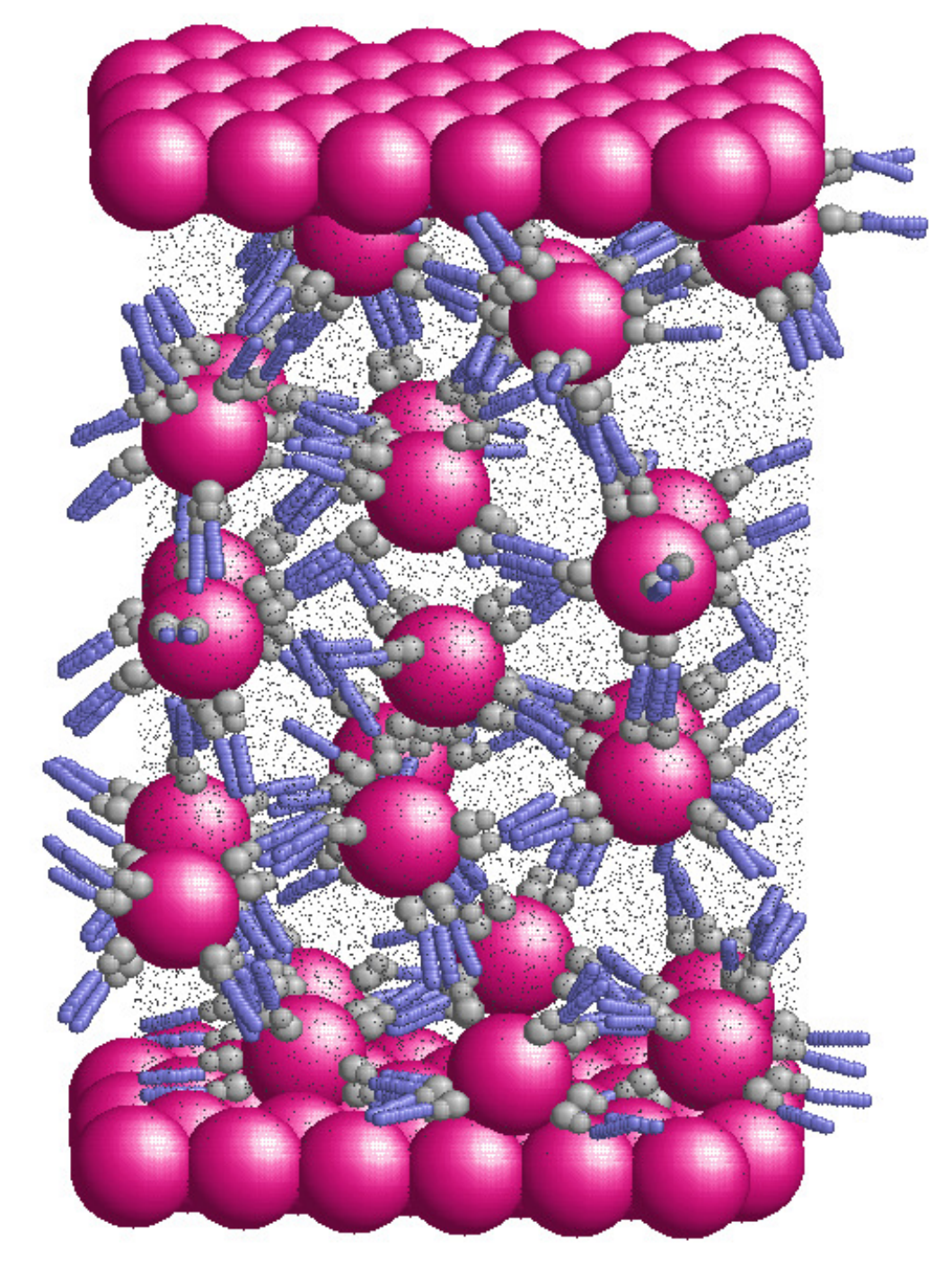}
\includegraphics[clip,angle=0,height=4cm]{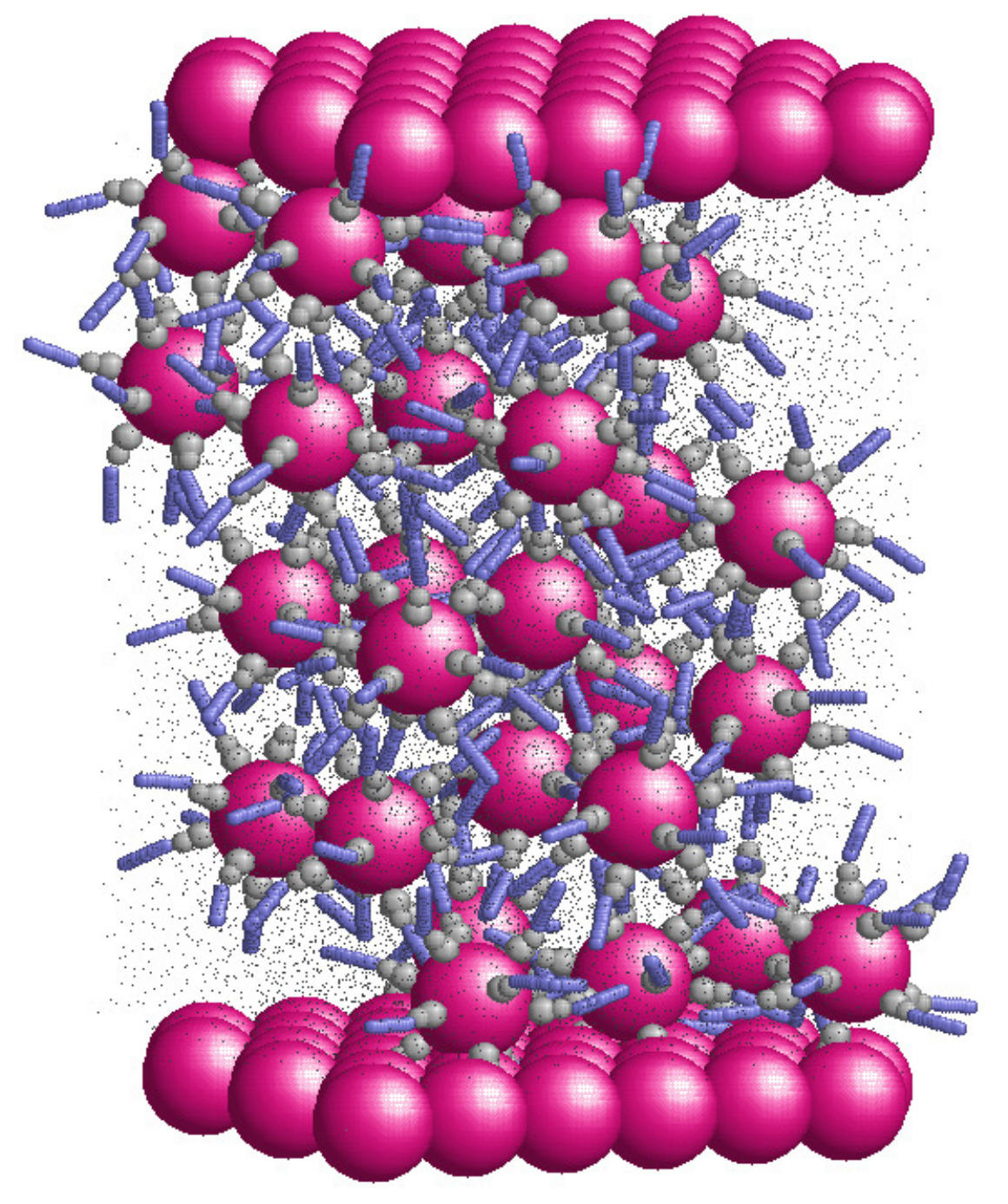}
\caption{\label{Snaps_Nm=27}Snapshots illustrating typical structure of the PGNPs gel formed for each patching
pattern introduced in Fig.~\ref{model}.}
\end{center}
\end{figure}

To study gelation in the solution of PGNPs of different patching pattern and density, we perform in
each case five gelation runs of $10\ns$ each.
Each production run is preceded by the equilibration run of the
same duration. In the latter, the attractive form (\ref{SAP}) is used for the LC-solvent interaction,
whereas the repulsive form (\ref{SRP}) is used for the LC-LC interaction. As the result, any previously
formed links between PGNPs are broken and the solution is forced into a colloid dispersion state.
The gelation run is initiated by switching the interaction potentials. Namely,
LC-solvent interaction is made strongly repulsive, via Eq.~(\ref{SRP}) with the maximum energy of
$U=140\cdot 10^{-20}\J$, whereas the LC-LC interaction is made attractive via 
the relation (\ref{SAP}).
Both effects of attraction between the LC beads and the solvophobicity of the ligands promote formation of the
ligand-ligand connections between the adjacent PGNPs. These connections shape a general structure of the
gel, as displayed in Fig.~\ref{Snaps_Nm=27} for all six patching patterns shown earlier in Fig.~\ref{model}.
A quick glance on these snapshots indicates prevalence of the linear chains for the ROD pattern,
more branched structures for the cases of TRI and QTR patterns and the tendency of DSC PGNP to form
two-dimensional sheets. The structures formed by the AXI and HDG patterns are highly branched.   

To quantify the differences in the gel structure formed for various patching patterns we treat it
as a network, where PGNPs serve as vertices and the connections between them via ligands -- as the links.
Namely, we register the $i$th and $j$th PGNPs as being linked if the LC beads $l$ of $i$th PGNP and
$m$ from $j$th PGNP exist, such that
\begin{equation}\label{crit}
|(\vhat{e}_l\cdot\vhat{e}_m)|>0.966,\;\;\;\mathrm{and}\;\;\; r_{lm}<1.25D,
\end{equation}
i.e. the beads $l$ and $m$ are almost collinear and arranged in a close side-to-side way. Based on
this criterion, we fill-in two arrays: the $\mathrm{Lnk}(i,j)$ and $\mathrm{Nlnk}(i,j)$. 
The element $(i,j)$ of the former array is equal to $1$ if at least one link between the $i$th and $j$th
PGNP exists and $0$ otherwise. The element $(i,j)$ of the latter array
is equal to the total number of links existing between the $i$th and $j$th PGNPs. By using the
$\mathrm{Lnk}(i,j)$ array, we build the list of the linked neighbors $\mathrm{Neigh}(i)$ for each $i$th PGNP
and keep track of their number in $\mathrm{Nn}(i)$.

After these preliminary steps, the network structure of a gel is studied. To this end we pick the $i$th seed PGNP
and identify all PGNPs linked to it. This subnetwork acquires the index $m$, $\mathrm{Subnet}(j)=m$, which is set
for the set of $\{j\}$ PGNPs that belong to it. The process is repeated starting from a new random
$i$th PGNP outside the subnetwork $s$ which seeds the subnetwork $s+1$. The stepwise algorithm is
as follows:
\begin{enumerate}
\item initiate $\mathrm{Subnet}(i):=0$ for all $i$, set first subnet index $m=0$;
\item pick $i$th PGNP with $\mathrm{Subnet}(i)=0$ randomly;
\item set subnet index $m:=m+1$, assign $\mathrm{Subnet}(i):=m$, mark $i$ as the newcomer to subnet $m$;
\item loop over the $\mathrm{Nn(i)}$ elements of the array $\mathrm{Neigh}(i)$ for all newcomers to
      the $m$th subnet and add them to the newcomer list;
\item repeat steps 4-5 until no newcomers appear;
\item go to step 2.
\end{enumerate}
Alternatively, the Hoshen and Koppelmann\cite{Hoshen1976} algorithm can also be used.

\begin{figure}
\begin{center}
\includegraphics[clip,angle=270,width=4cm]{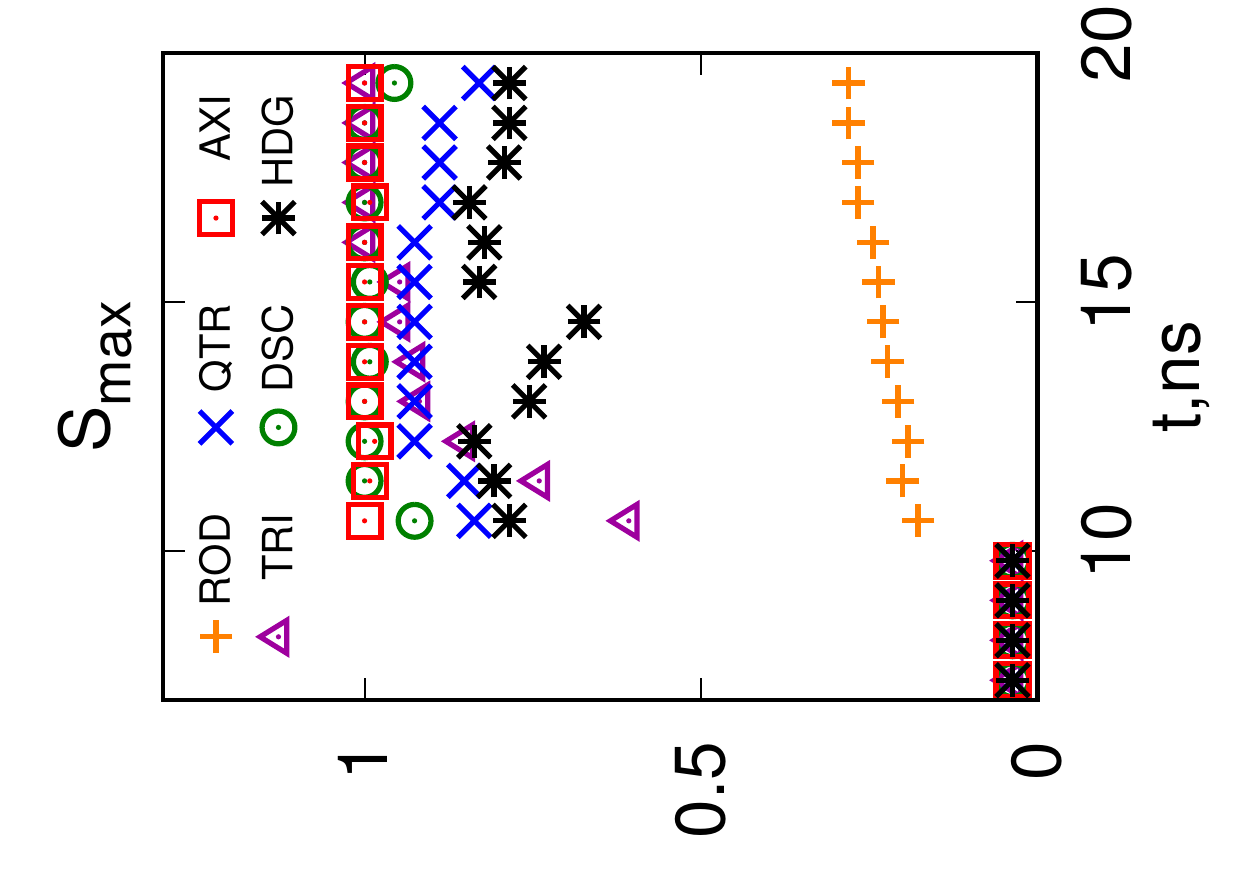}
\includegraphics[clip,angle=270,width=4cm]{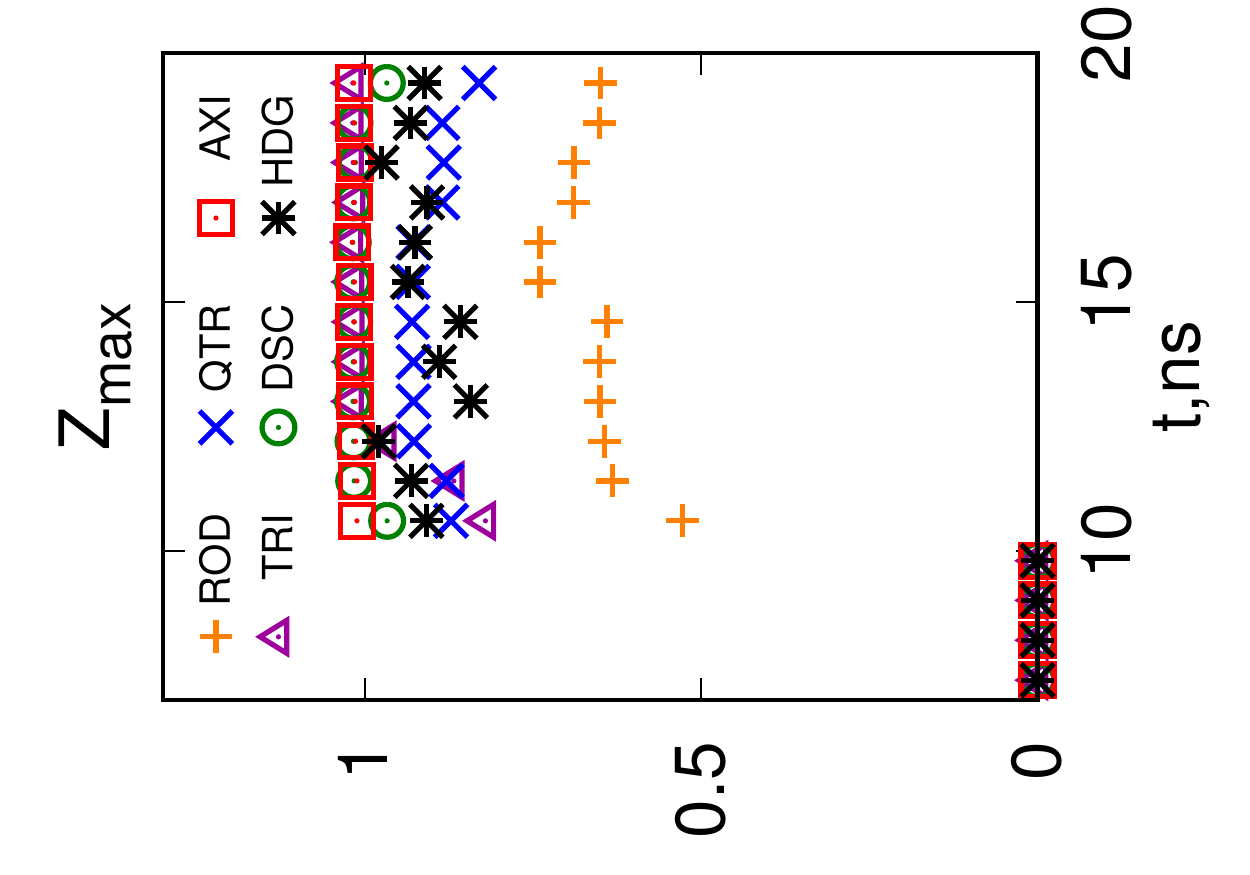}
\caption{\label{Sm_Zm_evol}Time evolution for the reduced size $S\idx{max}$ and intrapore span $Z\idx{max}$
of the largest subnet averaged over five gelation runs. The start of gelation occurs at $t=10\ns$ after
the equilibration run. Results for the solution of $N\idx{mol}=27$ PGNPs are shown for the six different types of
PGNPs nomenclatured according to Fig.~\ref{model}.}
\end{center}
\end{figure}
At the end of the subnets identification, each $m$th subnet acquires its size $S(m)$ that is equal to the number of
PGNPs that it contains. Our chief interest is in the largest of subnets, so-called ``giant component'', for which we evaluate
the reduced size $S\idx{max}$ and the intrapore span $Z\idx{max}$ defined as
\begin{equation}\label{clust_char}
S\idx{max}=\frac{S(m)|\idx{max}}{N\idx{mol}},\;\;
Z\idx{max}=\frac{\mathrm{span\; in\; Z}|\idx{max}}{L_z - 2\sigma\cos(\pi/6)},
\end{equation}
For a single network case one has $S\idx{max}=1$, whereas in a dispersed state of disjoint PGNPs
$S\idx{max}=1/N\idx{mol}$. The intrapore span $Z\idx{max}$ is defined as the $Z$-component of the reduced
separation between the two GNPs nearest to the $Z=0$ and $Z=L_z$ walls, respectively. 
The scaling factor $L_z-2\sigma\cos(\pi/6)$ is the maximum possible span, which takes into account
the near-wall regions inaccessible for the GNPs due to the presence of a layer of frozen particles with
the diameter $\sigma$. The intrapore span characterizes the percolating properties of the maximum net,
which is the case if  $Z\idx{max}\approx 1$.

We will consider the time evolution of $S\idx{max}$ and $Z\idx{max}$ at a fixed concentration
of the PGNP given by their number $N\idx{mol}=27$. Both properties are averaged over five
gelation runs. The plots shown in Fig.~\ref{Sm_Zm_evol} differ in the patching pattern.
First of all, the ROD type of PGNPs demonstrates its inability to form neither a single network nor
an intrapore percolating cluster, as far as both $S\idx{max}$ and $V\idx{max}$ 
are far from reaching $1$.
This is explained by the one-dimensional symmetry of this patching pattern which favors
the formation of linear chains of PGNPs but not branched structures, see also Fig.~\ref{Snaps_Nm=27} (a).
The HDG PGNPs perform a bit better, but a single
network regime is not reached with $S\idx{max}$ value peaking at  $0.8$.  We explain this result
by a stiff uniform radial ``hedghog''-like architecture of this patching pattern which restricts the orientation
freedom of ligands and, therefore, reduces the probability of link formation, see also Fig.~\ref{Snaps_Nm=27} (f).
This is the case at least for
the given grafting density of ligands and the chosen strength of the interpatch bonded interaction $k'_a$
in Eq.~(\ref{Vb}). One may expect that an increase of the former and a decrease of the latter may
turn the HDG patching pattern into a better candidate for efficient gelation but this is beyond the current study.
One could refer to the other studies on the role of the ligands mobility.\cite{Baran2017a} The QTR patching pattern also
displays incomplete gelation with both $S\idx{max}$ and $Z\idx{max}$ topping at around $0.8-0.9$. Other
three patching patterns show complete gelation and interpore percolation feature with the AXI one being the
fastest in formation of a gel and the TRI -- the slowest one.

The maximum subnet size $S\idx{max}$ and the intrapore span $Z\idx{max}$ provide the simplest account of
the network properties only and are supplemented by the characteristics of their internal connectivity.
In particular, the rank of the vertex $k(i)$
provides the number of its links to the other vertices. The local clustering 
coefficient $c(i)$ is defined as the ratio between
the number of links formed between the linked neighbors of the $i$th vertex PGNP and the total number of
the pairs formed by them. When evaluating these, we use the array $\mathrm{lnk}(i,j)$,
thus we ignore possible multiple links between the PGNPs, hence
\begin{eqnarray}\label{k_c_ordinary}
k(i) &=& \sum_{j=1}^{\mathrm{Nn}(i)}\mathrm{Lnk}(i,j)=\mathrm{Nn}(i),\\
c(i) &=& \sum_{j=1}^{\mathrm{Nn}(i)-1}\sum_{l=j+1}^{Nn(i)}
                \frac{\mathrm{Lnk}(j,l)}{\mathrm{Nn}(i)[\mathrm{Nn}(i)-1]/2}.
\end{eqnarray}
These characteristics are averaged then over all PGNPs yielding
\begin{eqnarray}\label{k_c_aver}
K=\frac{1}{N\idx{mol}}\sum_{i=1}^{N\idx{mol}}k(i),~~ C=\frac{1}{N\idx{mol}}\sum_{i=1}^{N\idx{mol}}c(i)
\end{eqnarray}
\begin{figure}
\begin{center}
\includegraphics[clip,angle=270,width=4cm]{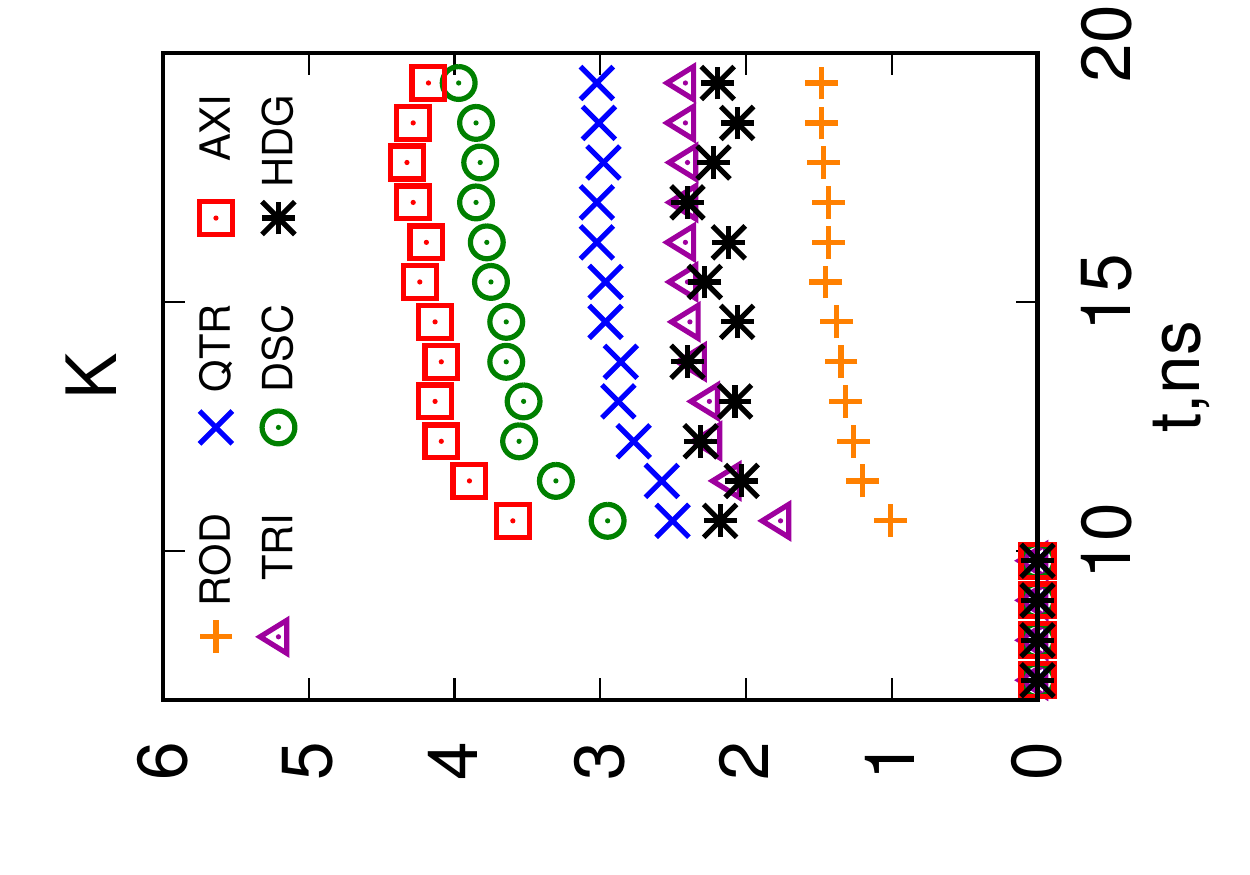}
\includegraphics[clip,angle=270,width=4cm]{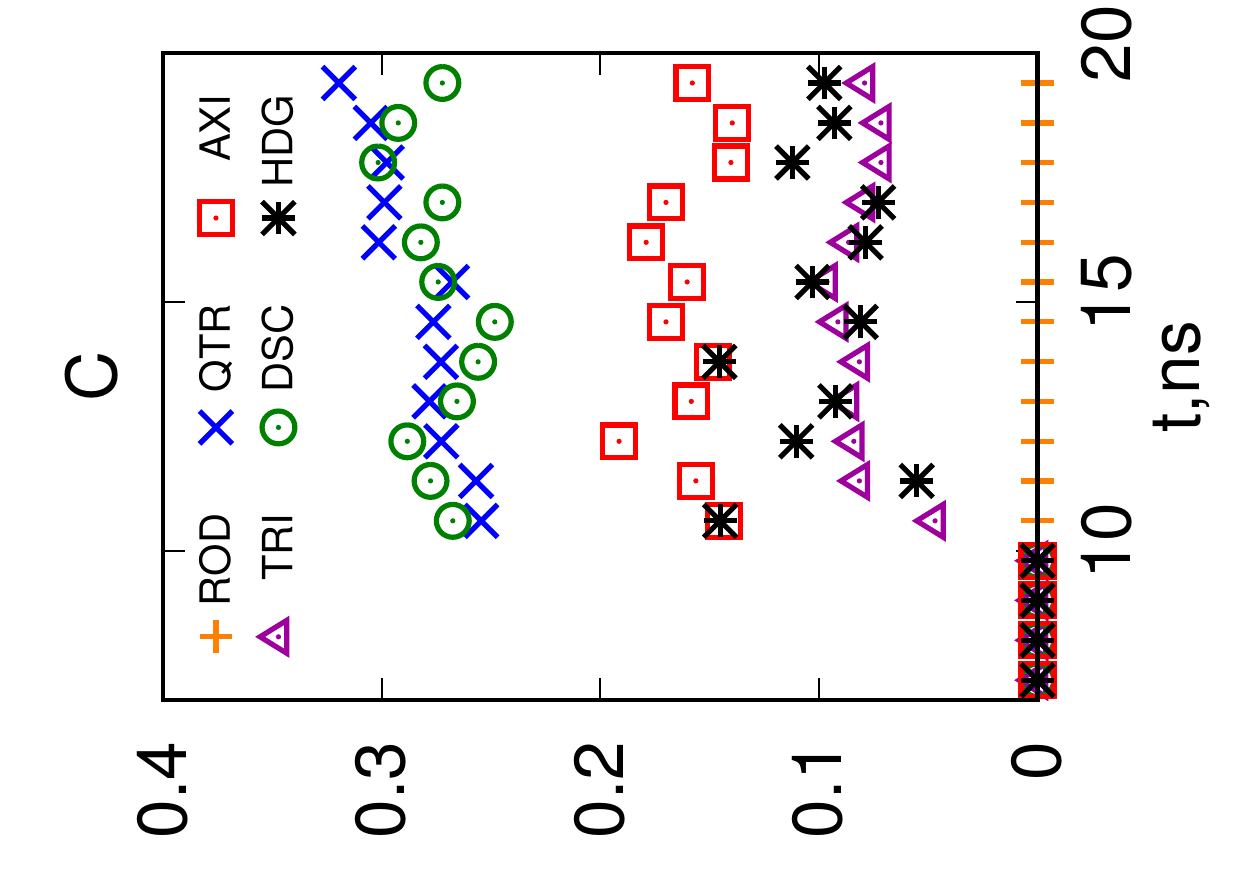}
\caption{\label{K_C_evol}The same as in Fig.~\ref{Sm_Zm_evol} but for the time evolution of the average rank
$K$ and of the local clustering coefficient $C$.}
\end{center}
\end{figure}
The evolution of the average rank $K$ and of the local clustering coefficient $C$ 
are given in Fig.~\ref{K_C_evol}. The DSC and AXI
PGNPs demonstrate the highest average rank of $K\approx 4$ indicating 
highly branched structures. For the case of the
QTR PGNP this value is lower, $K\approx 3$ and it drops further down 
to about $K\approx 2$ for the TRI and HDG,
indicating the dominance of liner fragments and weak branching of the latter two gels.
This correlates well with the ability
of the respective patching patterns to form a single percolating network, 
as discussed in relation to Fig.~\ref{Sm_Zm_evol}.
The DSC and QTR PGNPs produce the most locally clustered network with the value of $C$ approaching about $0.3$.
The value of $C$ for the case of AXI and TRI PGNPs is lower and is close to $0.15$ and $0.09$, respectively.

\begin{figure}
\begin{center}
\includegraphics[clip,angle=0,width=6cm]{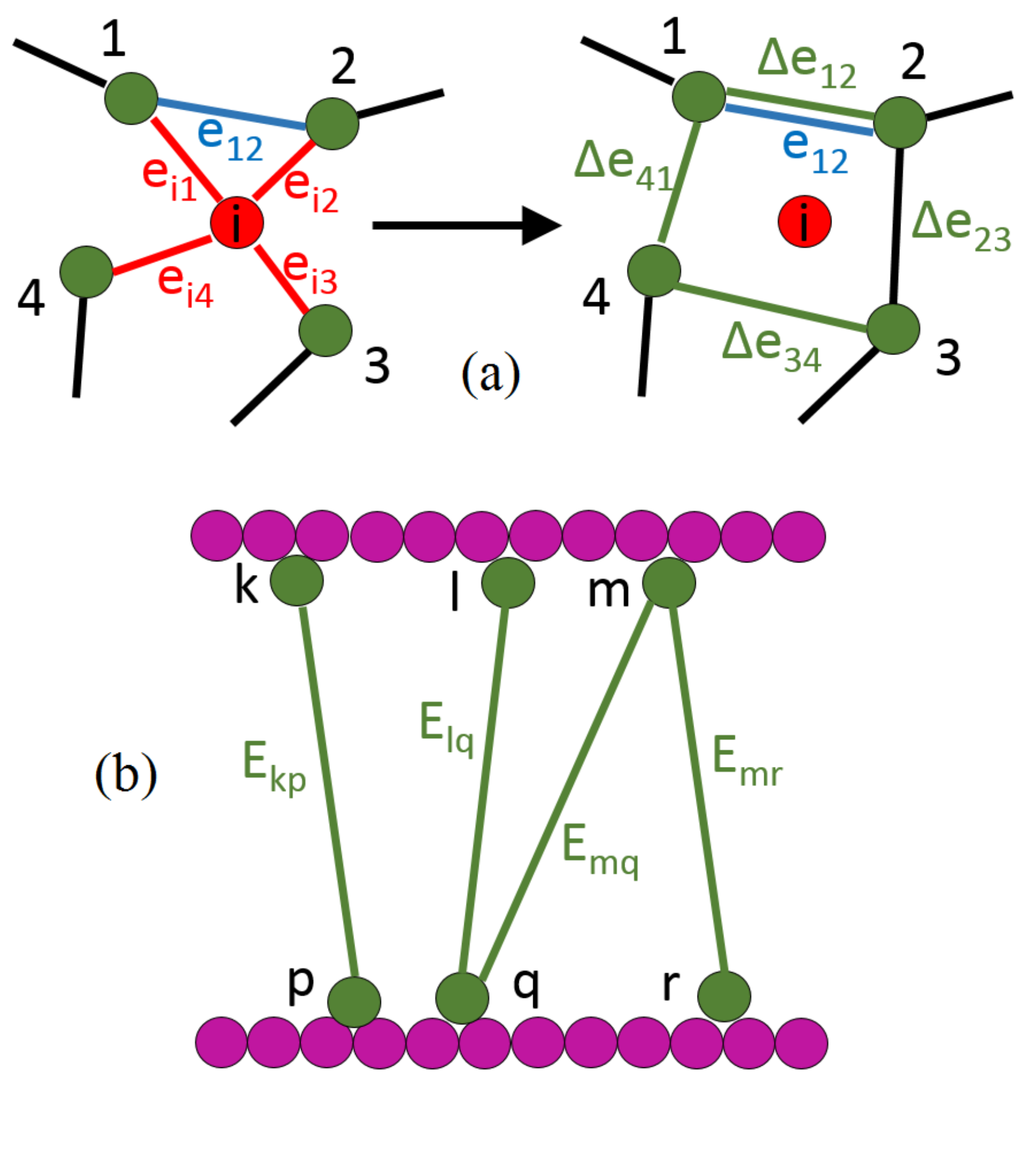}
\caption{\label{elim}Illustration of the vertex elimination algorithm for evaluation of the effective
spring constant of the network with respect to two walls.}
\end{center}
\end{figure}
Both characteristics, $K$ and $C$, affect mechanical robustness of the network, the property that is important for the
practical application of the GNPs gel for catalysis. It can also be addressed in a more direct way, by evaluating
elastic properties of the network of PGNPs in the direction perpendicular to the walls.
To this end, we
assume the network to be pinned to the walls by the PGNPs adsorbed on the walls, whereas each link between the
$i$th and $j$th PGNPs (where at least one of them is not a pinning PGNP) acquires the Hookean spring
constant of $e_{ij}=1$. Then the effective spring constant $E$ is evaluated for a network of springs between the
pinning PGNPs. As far as the spring constant for the parallel and serial connection of two springs follows
that for the conductance of electric resistors, the same Kirkhoff's rule based calculations can be employed.
These are can done efficiently by the vertex elimination algorithms following Ref.~\cite{Fogelholm1980}
\begin{enumerate}
\item choose vertex $i$ which is not a grafting point;
\item evaluate the sum of spring constants towards its neighbors $E_i=\sum_{k=1}^{Nn(i)}e_{ik}$;
\item add contribution $\Delta e_{kl}=e_{ik}e_{il}/E_i$ to the spring constant of each pair
      $\{k,l\}$ of its neighbors, see Fig.~\ref{elim} (a);
\item eliminate vertex $i$ from the network;
\item repeat steps 1-4 until only grafting points are left;
\item evaluate total wall-to-wall spring constant as $E=\sum_{p,q}E_{pq}$ over all pairs of grafting point $\{p,q\}$
      that belong to opposite walls, see Fig.~\ref{elim} (b).
\end{enumerate}
\begin{figure}
\begin{center}
\includegraphics[clip,angle=270,width=4cm]{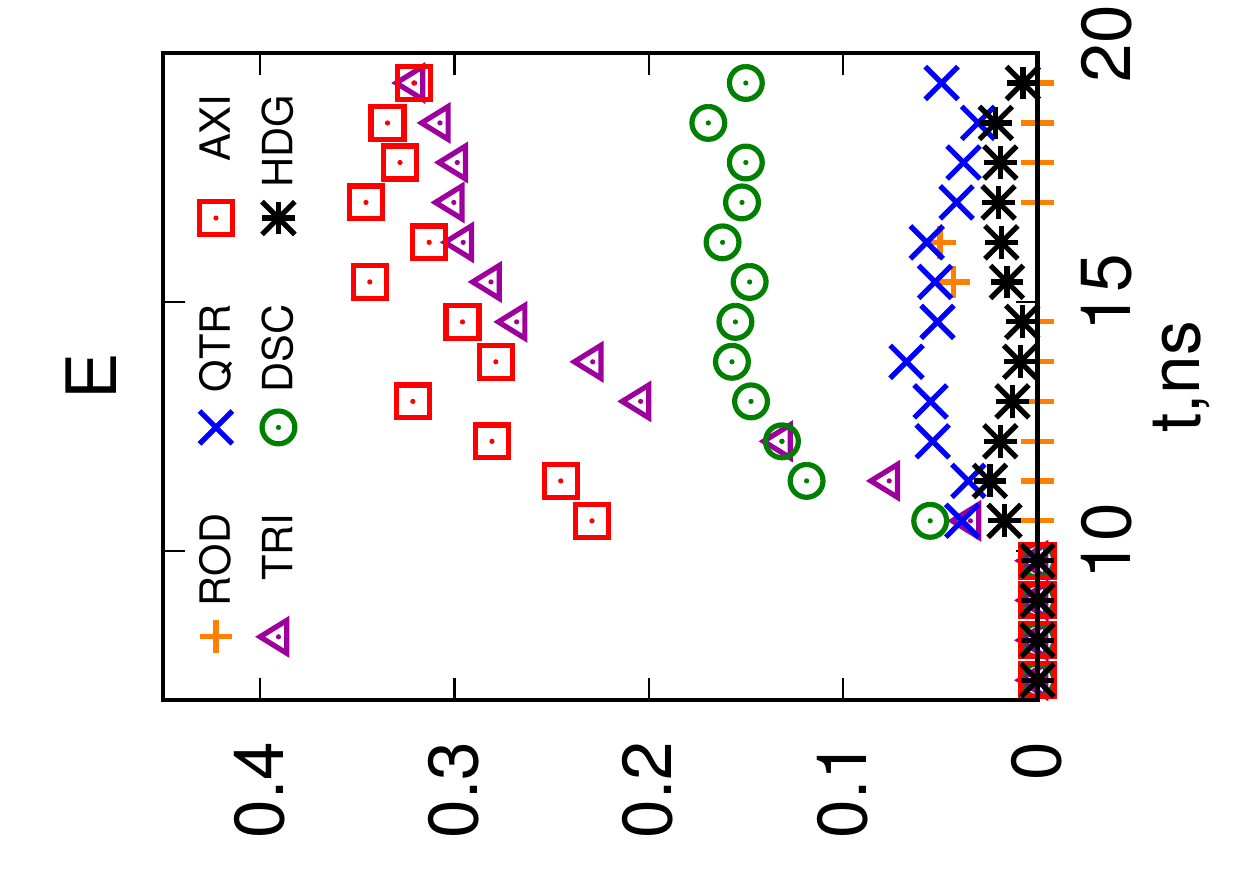}
\caption{\label{E_evol}The same as in Fig.~\ref{Sm_Zm_evol} but for the time evolution of the
effective spring constant $E$ of the network evaluated between two walls.}
\end{center}
\end{figure}
Evolution of the network spring constant $E$ is shown in Fig.~\ref{E_evol} for the case of $N\idx{mol}=27$
PGNPs and various patching patterns. It indicates that for this particular case the AXI and TRI patterns
produce the networks with the highest values of $E\approx 0.3-0.35$
whereas the values of $E$ for other patterns are at least twice lower.

Based on the analysis of a gel network at $N\idx{mol}=27$ performed above, one can summarize the following.
Out of six patching patterns only three, AXI, DSC and TRI, form a single network and intrapore percolating structure,
with the QTR being just short of achieving the same. The remaining, ROD and HDG patterns, are unable
to form a single network. The AXI PGNPs form a highly branched weakly clustered gel, the DSC one --  almost the
same highly branched but highly clustered network, both on a time scale of up to $1\ns$. The TRI network is formed
relatively slowly, on a time scale of about $5\ns$ and the resulting network is both weakly branched and weakly clustered.
The effective spring constant of the network with respect to the walls is the highest for the cases of AXI and TRI
patterns, both characterized by a low local clustering and is low for the case of DSC pattern characterized
by a low local cluster coefficient. This has a simple intuitive explanation -- if most of links are spent for
connecting PGNPs locally, then the remaining number of them is insufficient to ensure strong global
interconnectivity of a network needed for its high effective spring constant. The analysis is extended to the
broad interval of the concentration of PGNPs in Sec.~\ref{IV}.

\section{\label{IV}The properties of a gel network in a broad interval of concentrations of PGNPs}

Let us consider now the properties of the PGNPs related to its potential application as a catalyst. The effectiveness
of such heterogeneous catalysis depends on a number of factors. The first factor 
is the total surface area of GNPs
accessible for the reagents. Let us assume that that the chemisorbed reagent occupies approximately the area
of $\pi r_r^2$ on the sphere of radius $r_{GNP}+r_s$, where $r_{GNP}$
is the radius of GNP and $r_r$ is the effective steric radius of the group of atoms of a reagent that are the
closest to the surface of GNP. The radius of a GNP used in this study is $r_{GNP}=0.107\nm$, whereas the
value of $r$ equal to the radius of a solvent bead, $r_r=0.23\nm$, can be used as a first approximation.
Within these assumptions, each GNP can accommodate approximately $104$ molecules of a reagent,  after one takes into account
the surface area already occupied by grafted ligands. The total grafting capacity of a gel is then equal to
the number of reagents it can accommodate: $N_c=104N_{mol}$.

\begin{figure}
\begin{center}
\includegraphics[clip,angle=270,width=6cm]{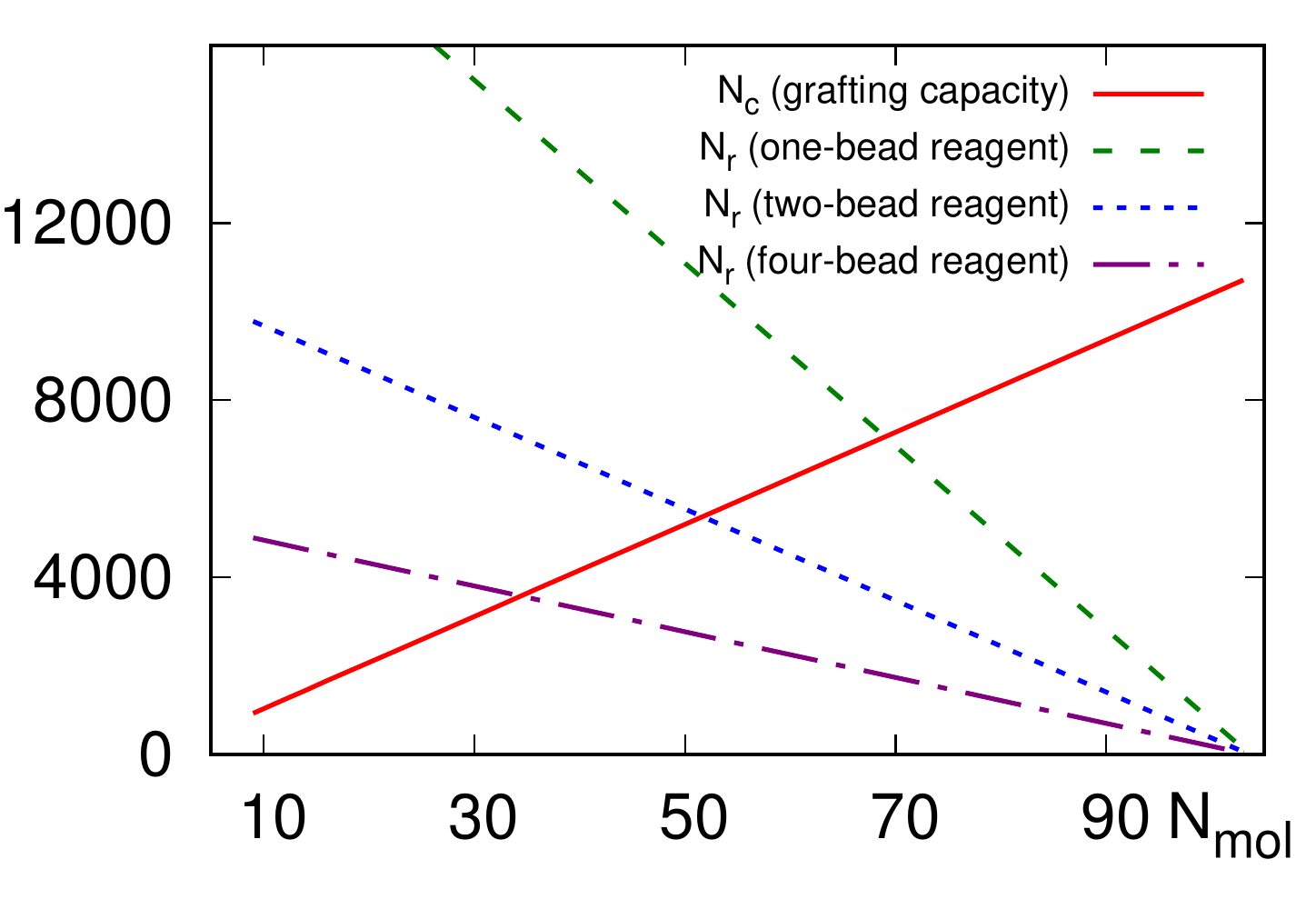}
\caption{\label{N_r}Estimate for the effectiveness of catalysis based on the grafting capacitance
$N_c$ of the total surface area of PGNPs and the number $N_r$ of available reagent molecules
at given concentration of PGNPs $N\idx{mol}$, see text for details.}
\end{center}
\end{figure}
The second factor is the availability of the required number of reagents' molecules. 
As far as the density of the solution
is constant, the increase of $N_{mol}$ leads to the decrease of the number of reagents' molecules, $N_r$,
the exact number depends on the excluded volume of each reagent. The latter can range in a wide interval,
as demonstrated in Refs.~\cite{Raffa2008,Wei2010,Wu2015} To provide some estimates we assume that
all the interior of a gel is occupied by a reagent only and a single reagent molecule has excluded volume
equivalent to that of $n$ solvent beads. In this case one has
$N_r=(V - N\idx{mol}V\idx{mol})/(nv_s)$, where $V\idx{mol}$ is an approximate excluded volume
of a single PGNP and $v_s$ is the volume of a single solvent bead. 
The dependencies of $N_c$ and $N_r$
on $N\idx{mol}$ provided for $n=1,2,4$ are displayed in Fig.~\ref{N_r}. It is obvious that the catalysis
takes place within the intersection of the areas below the $N_c$ line and of respective $N_r$ line. The top
of this intersection area has a triangular form indicating that the maximum efficiency of catalysis is to be looked
for within the interval of $30<N\idx{mol}<70$.

The third factor that affects the effectiveness of the catalysis is related to the diffusion of reagents: 
first to the surface of the nearest GNPs to chemisorbed there,
and then away of it after the reaction took place. This
effect depends strongly  on the porosity of the gel, as well as on the molecular 
topology of the reagents and requires
specially tailored simulations that fall beyond the scope of a current study. One would expect that with the
increase of $N\idx{mol}$ the diffusion of reagents will slow down due to both narrowing internal pores
and getting their interconnected structure more complicated. This would further reduce the effectiveness of
the network with higher values of $N\idx{mol}$ for catalytic applications.

\begin{figure}
\begin{center}
\includegraphics[clip,angle=270,width=6cm]{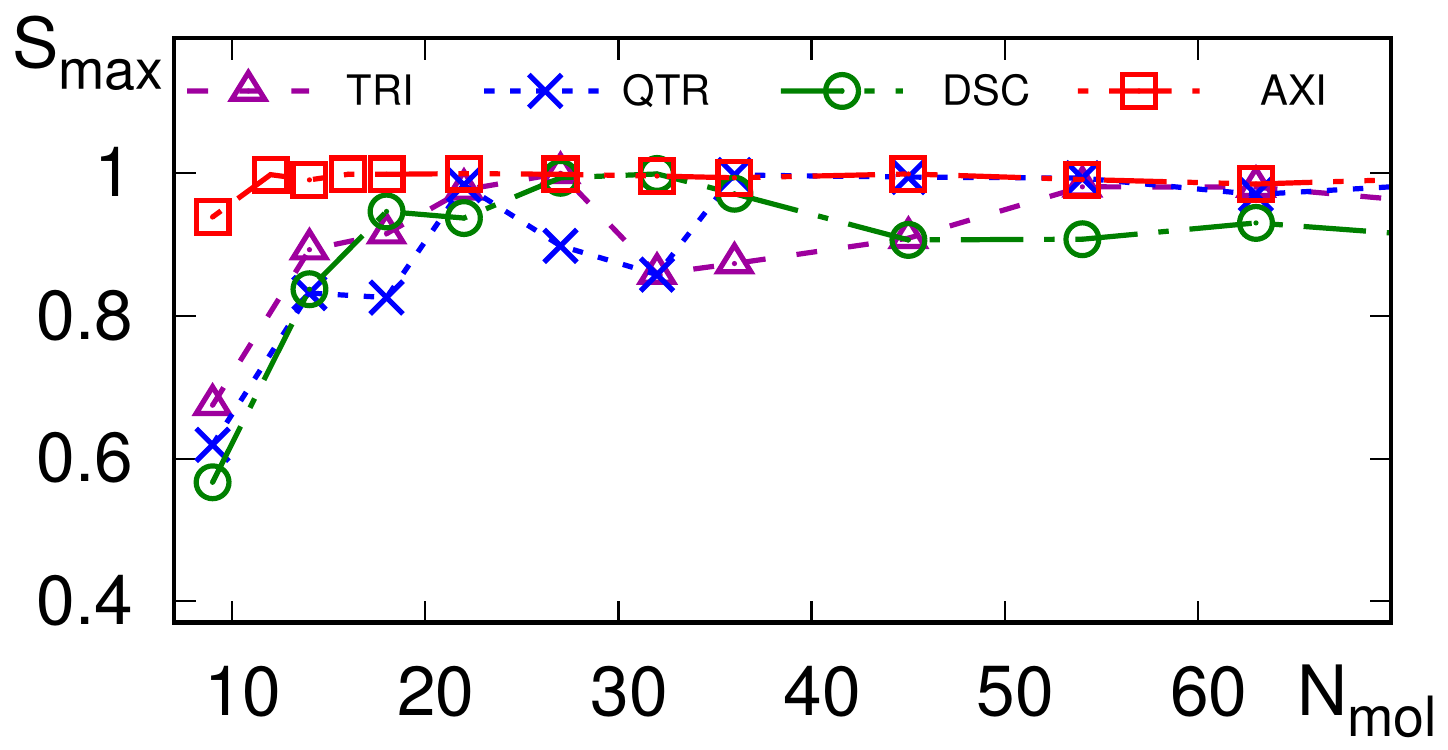}\\
\includegraphics[clip,angle=270,width=6cm]{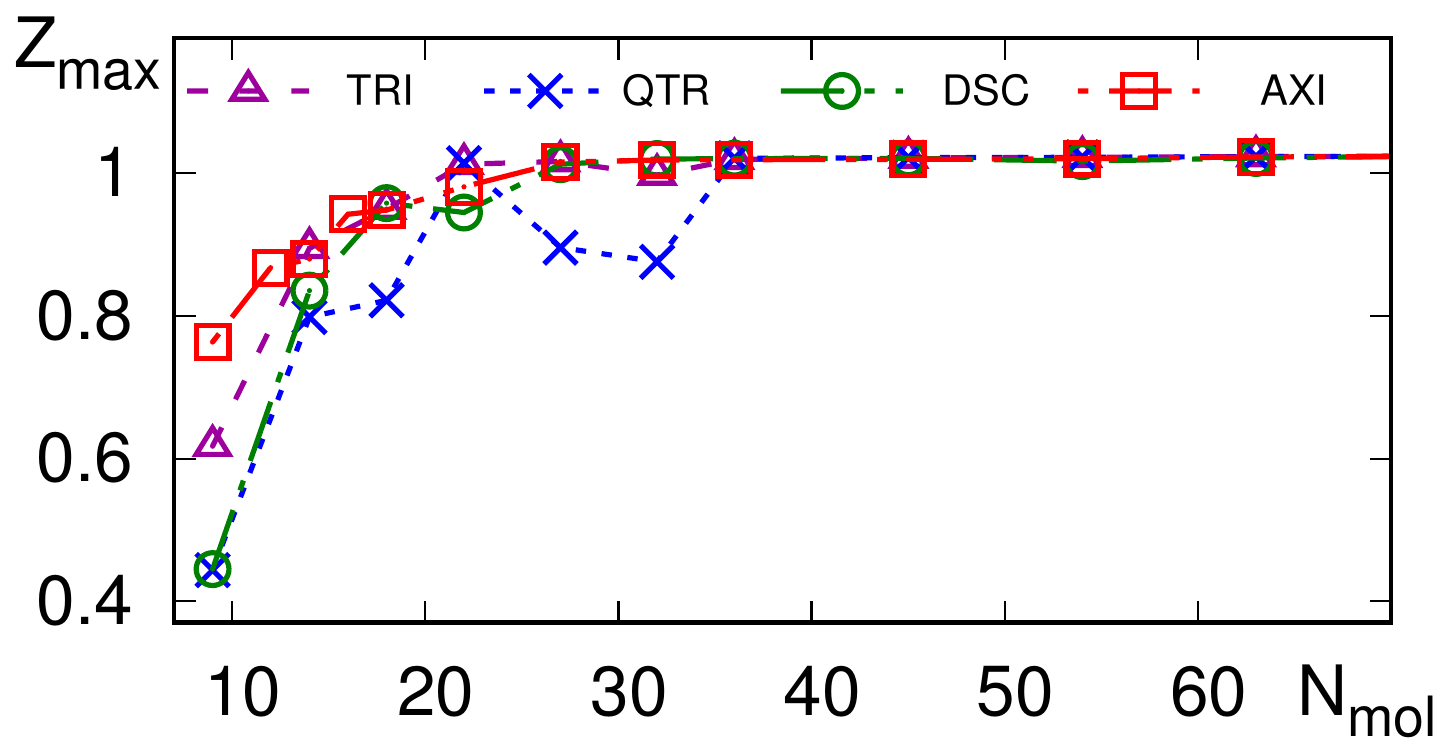}
\caption{\label{Sm_Zm_Nmol}Average reduced size $S\idx{max}$ and intrapore span $Z\idx{max}$
of the largest subnet vs the number $N\idx{mol}$ of PGNPs.}
\end{center}
\end{figure}
Based on these simple calculations, we restrict our attention to the interval of $N\idx{mol}<70$.
For each selected value of $N\idx{mol}$ and for each of the four patching patterns, TRI, QTR, DSC and AXI,
we performed five gelation runs. The properties of interest, $S\idx{max}$, $Z\idx{max}$, $K$, $C$ and $E$, are
averaged over the last $3\ns$ of all five gelation runs. The dependencies of $S\idx{max}$ and $Z\idx{max}$
vs  $N\idx{mol}$ are shown in Fig.~\ref{Sm_Zm_Nmol}. These 
indicate that the AXI pattern is the most robust in terms of the formation
of a single network and wall-to-wall percolating structure in a wide interval of $N\idx{mol}$.

\begin{figure}
\begin{center}
(a)\hspace{12em}(b)\vspace{-1.5em}\\
\includegraphics[clip,angle=270,width=4cm]{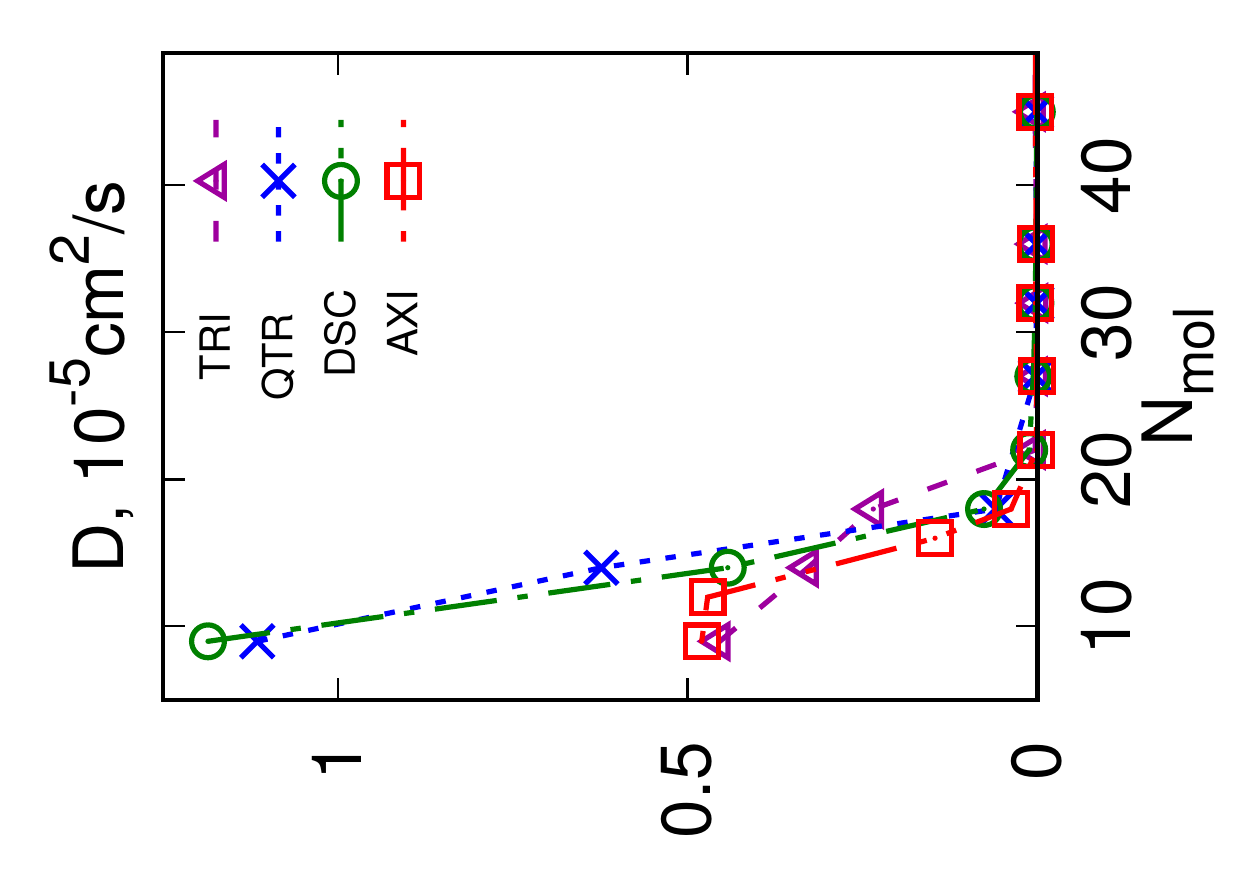}
\includegraphics[clip,angle=270,width=4cm]{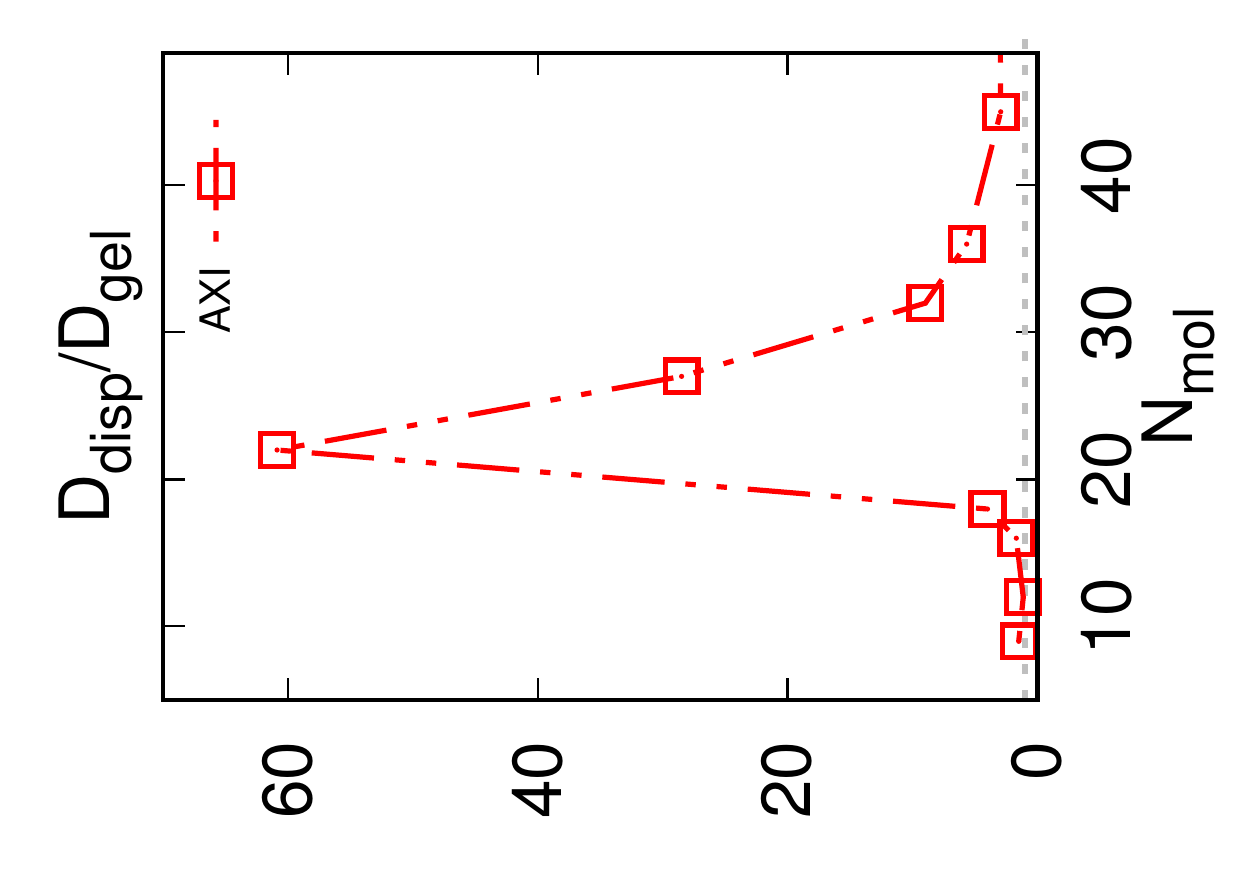}
\caption{\label{Diff_coef}(a) Changes of the diffusion coefficient $D$ of the GNPs in a gel state
upon increase of the number of PGNPs $N\idx{mol}$ shown for each of four patching patterns TRI, QTR,
DSC and AXI. (b) The same shown 
for the AXI pattern only (marked as ``gel'') and compared with the diffusion
coefficient of the same pattern in the dispersed state evaluated during the equilibration runs (marked as ``disp'').}
\end{center}
\end{figure}
As expected, gelation has a profound effect on diffusivity of the PGNPs. The diffusion coefficient $D$ is estimated for
the GNPs (the central bead of each PGNP) from the linear subparts of their mean square displacements vs. time
using Einstein's relation
\begin{equation}\label{Eins}
\langle\vvec{r}_i(t)-\vvec{r}_i(0)\rangle_{\mathrm{GNP},\mathrm{runs}}^2  = 6Dt,
\hspace{2em}{t\in[t\idx{min},t\idx{max}]},
\end{equation}
where averaging is performed over all GNPs and over five gelation runs performed for each combination
of the $N\idx{mol}$ and patching pattern. We found the shape of the mean square displacement to be
highly linear within the interval given by $t\idx{min}=4\ns$ and $t\idx{max}=10\ns$.
The results for the estimated diffusion coefficient are shown in Fig.~\ref{Diff_coef} (a) indicating sharp
decrease of $D$ at $N\idx{mol}<18$ followed by the the values of $D\approx 0$ at $N\idx{mol}>20$.
One should note the presence of two factors here -- one is related to the general decrease of diffusivity of
the solution with the increase of $N\idx{mol}$ even in a dispersed state, 
another -- is the pure effect of gelation.
To separate both we performed a comparison between the values of $D$ in a dispersed state
with that in a gel state for the case of AXI patching pattern. Dispersed state is mimicked by the equilibration
runs where the PGNPs are disjointed due to the appropriate tuning of the interaction potentials, as discussed
in Sec.~\ref{II}. The ratio of the two is shown in Fig.~\ref{Diff_coef} (b) indicating the most profound
effect of gelation to take place at $20<N\idx{mol}<30$, where the diffusion coefficient decreases up to $60$
times as compared to the dispersed state. 

\begin{figure}
\begin{center}
\includegraphics[clip,angle=270,width=6cm]{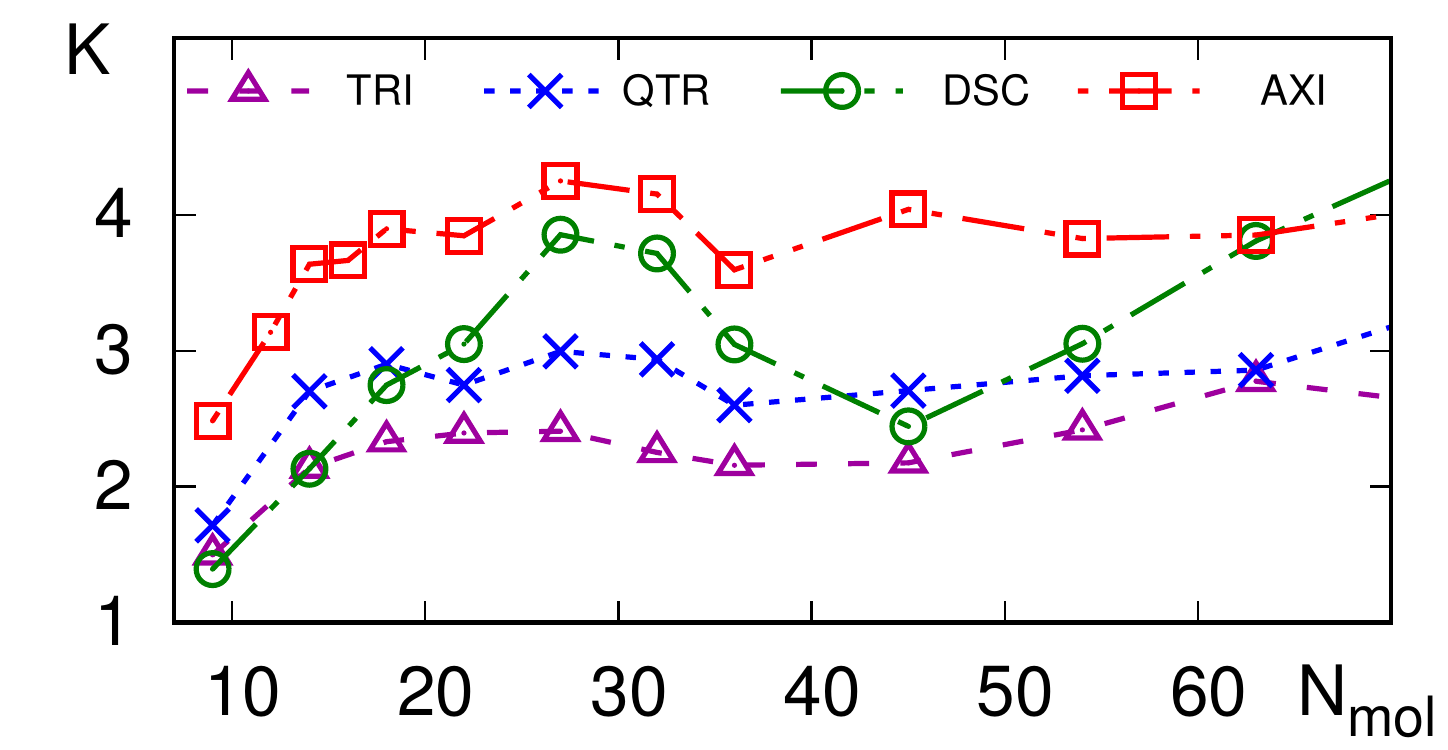}\\
\includegraphics[clip,angle=270,width=6cm]{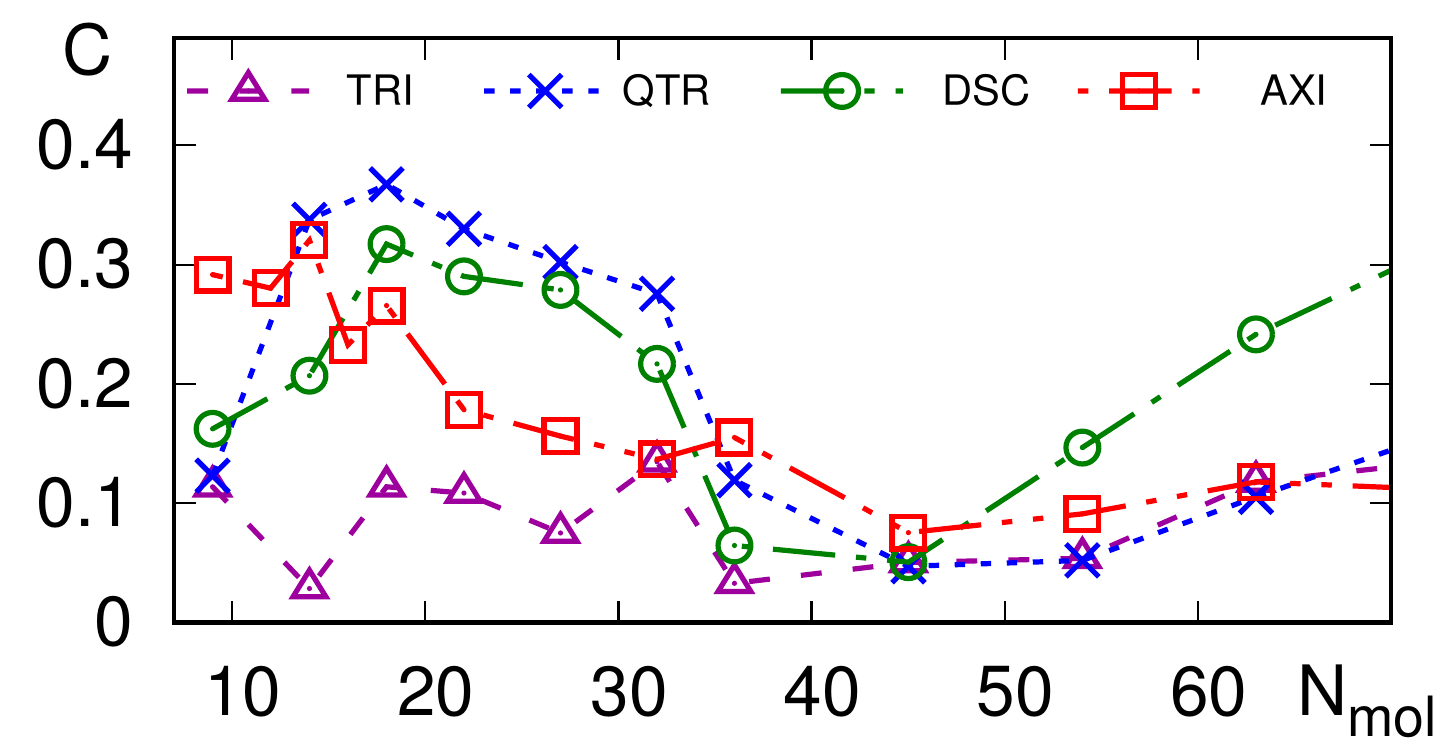}
\caption{\label{K_C_Nmol}Average rank $K$ and the local clustering coefficient $C$
of the largest subnet vs the number $N\idx{mol}$ of PGNPs for four patching patterns
indicated in the figure.}
\end{center}
\end{figure}
The plots for the dependencies of the average rank $K$ and the local clustering coefficient $C$
vs $N\idx{mol}$ are shown in Fig.~\ref{K_C_Nmol}. We conclude from these plots that the AXI
patching pattern is characterized by the highest values for the average rank $K\approx 4$
and the flattest dependence of this characteristic on $N\idx{mol}$. Similar values for $K$ are
approached only for the DSC pattern in a narrow intervals of $25<N\idx{mol}<30$ and
$N\idx{mol}\approx 60$. The local clustering coefficient $C$ shows a pronounced hill-like
shape at $9<N\idx{mol}<35$ with the lowest values at $35<N\idx{mol}<50$ and with its
values increasing again at $N\idx{mol}>50$.

\begin{figure}
\begin{center}
\includegraphics[clip,angle=270,width=6cm]{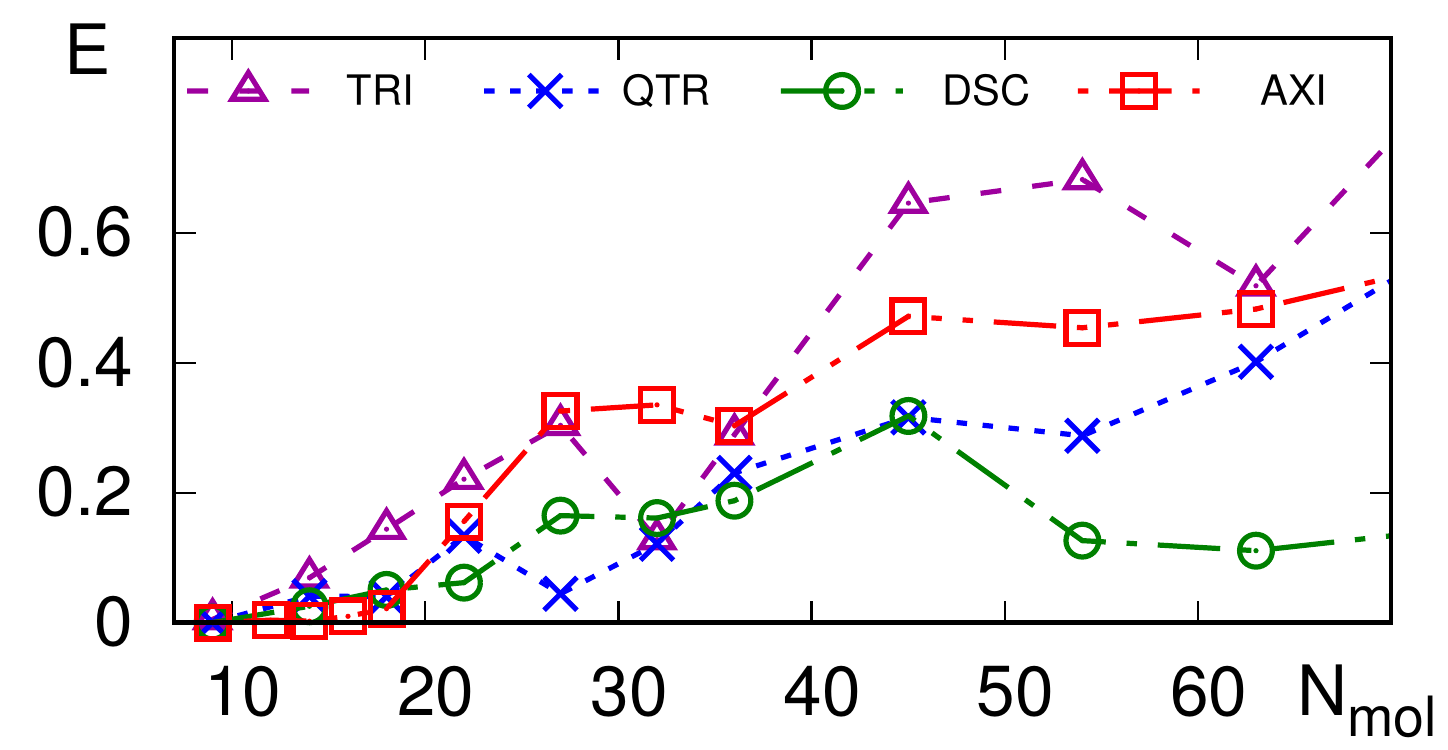}
\caption{\label{E_Nmol}Effective spring constant of the gel network $E$ vs the number
$N\idx{mol}$ of PGNPs for four patching patterns indicated in the figure.}
\end{center}
\end{figure}
The main property of interest is, in our view, the effective spring constant $E$ of the network
which characterizes its mechanical robustness and is defined in Sec.~\ref{III}. The values for $E$
vs $N\idx{mol}$ are displayed in Fig.~\ref{E_Nmol} for four patching patterns TRI, QTR, DSC
and AXI. The plot indicates that two patterns, AXI and TRI demonstrate the highest values of
$E$ in a whole interval of interest $9<N\idx{mol}<70$. Let us note that at $N\idx{mol}>20$
these two patterns are characterized by rather different values of the average rank: $K\approx 4$
and $K\approx 2.4$, respectively, but their clustering coefficient $C$ is rather similar,
$C=0.05-0.15$ as compared to the values for two other patterns. This indicates that the
observation made in Sec.~\ref{III} concerning the inverse proportionality of the effective spring
constant of the network and its clustering coefficient found at $N\idx{mol}=27$, holds in a broad interval
of $N\idx{mol}$ values. 

\begin{figure}
\begin{center}
(a)\vspace{-1.5em}\\
\includegraphics[clip,angle=270,width=6cm]{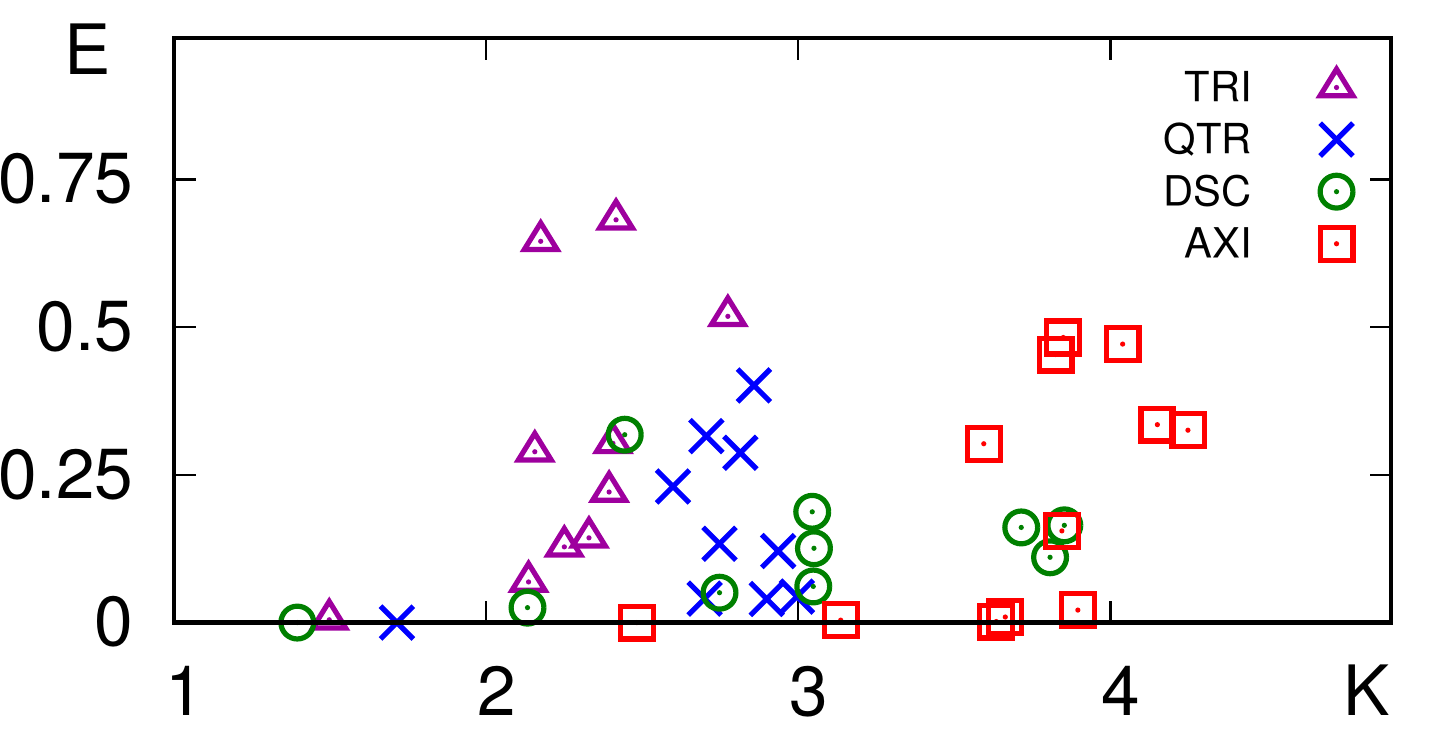}\\
(b)\vspace{-1.5em}\\
\includegraphics[clip,angle=270,width=6cm]{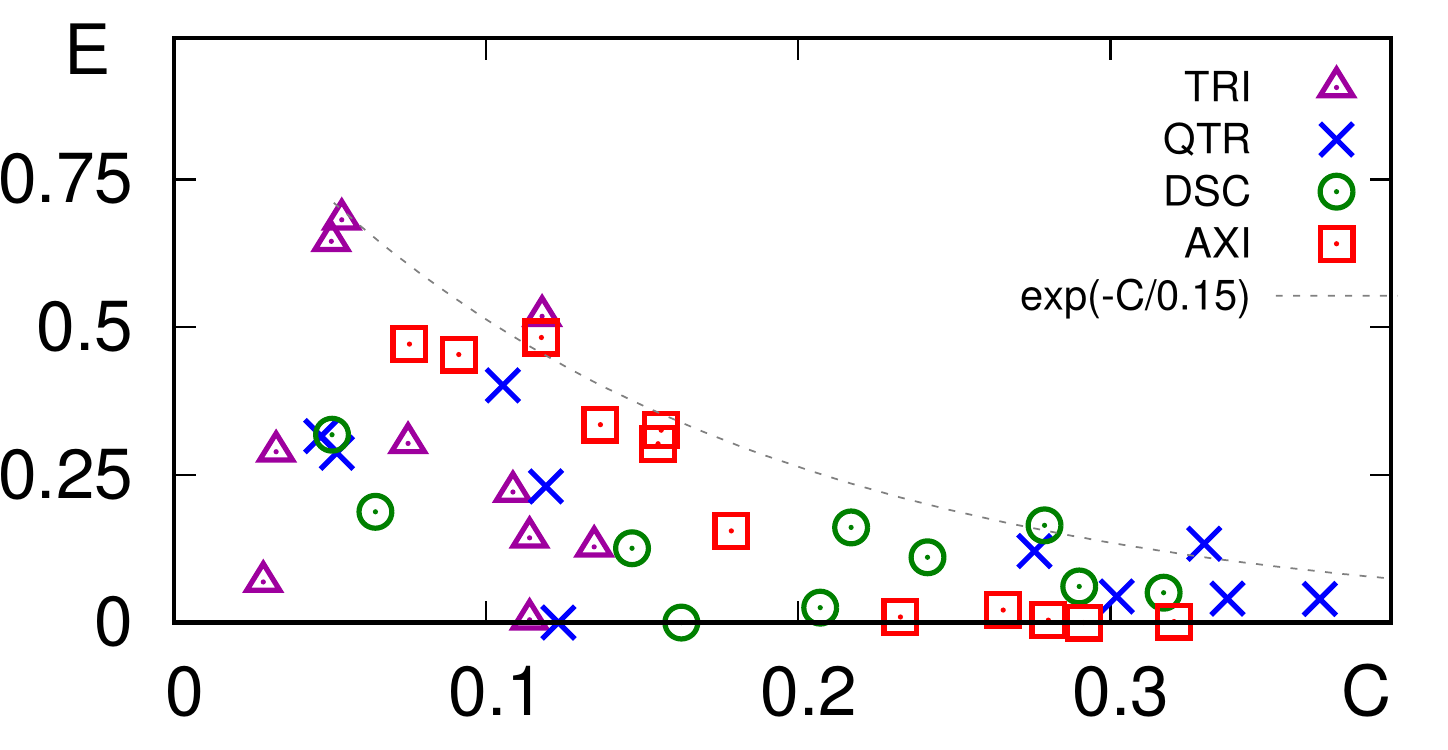}
\caption{\label{E_maps}(a) Scattered plot for the effective spring constant of the network $E$
vs its average rank $K$. (b) The same for the $E$ vs the local clustering coefficient $C$, dashed
curve is the guide for eyes for the upper boundary of the region with scattered points. Various
patching patterns are indicated in both plots.}
\end{center}
\end{figure}
To study the correlation between the effective spring constant $E$ vs average rank $K$ and $E$ vs local
clustering coefficient $C$ directly, we built the scattered plots for these respective pairs as shown in
Fig.~\ref{E_maps} (a) and (b) respectively. Plot (a) of this figure shows that the points that belong to
the same patching pattern are chiefly distributed as a bunches of columns with similar values of $K$.
The maximum values for $E$ are achieved at specific, pattern dependent, interval of $K$, e.g.
$K\approx 2.2$ for the TRI pattern, and $K\approx 4$ for the AXI pattern. This indicates no direct correlation
between the $E$ and $K$ values independent on the patching pattern. Fig.~\ref{E_maps} (b) shows quite
different result for the correlation between the $E$ and $C$ values. Namely, the available data points
fill-in the distinct area in a $\{K,E\}$ plane, which looks to be top bounded by an exponential curve.
This indicates that the increase of the local clustering coefficient restricts the maximum possible
value for the effective spring constant of the network. Therefore, the networks with lower $C$
are preferable for formation of mechanically robust gel with high values of the effective spring
constant $E$.

\begin{figure}
\begin{center}
(a)\vspace{-1.5em}\\
\includegraphics[clip,angle=270,width=6cm]{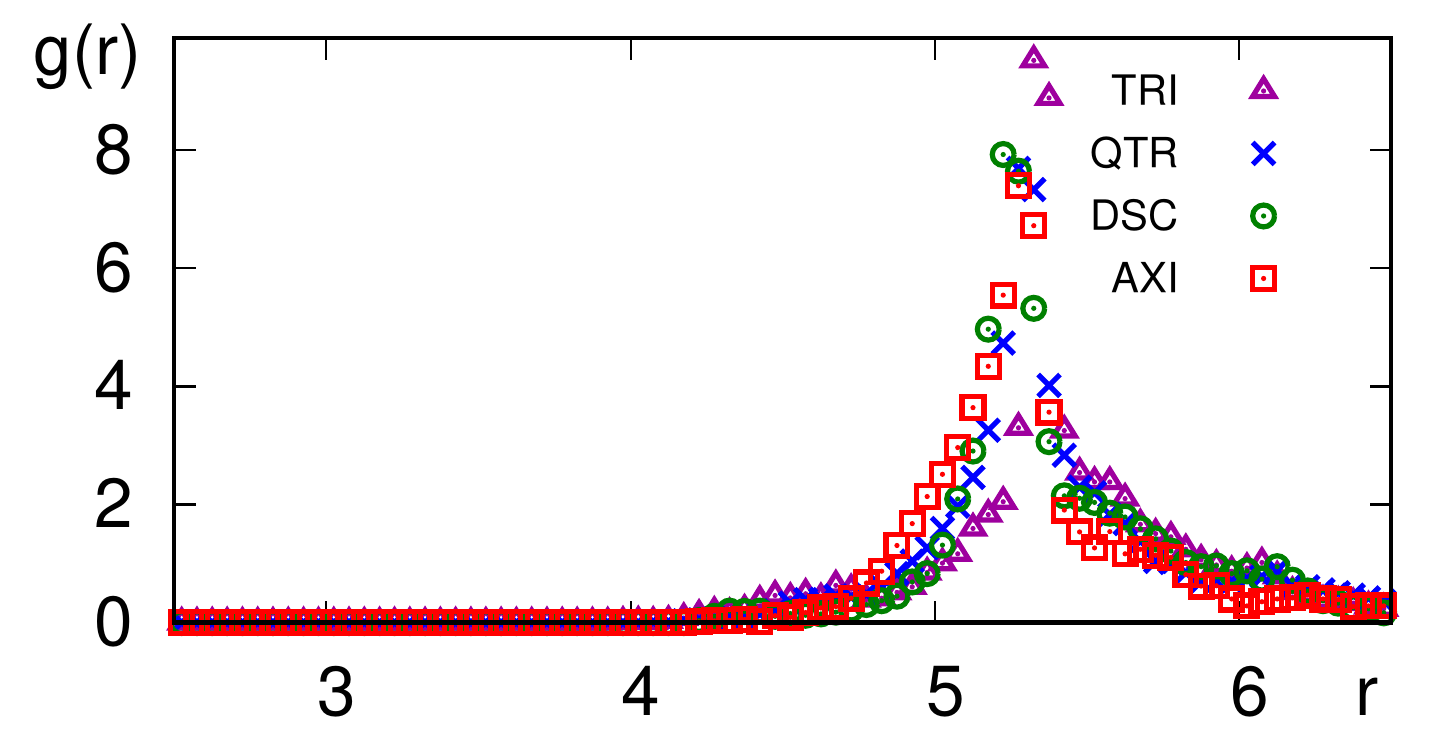}\\
(b)\vspace{-1.5em}\\
\includegraphics[clip,angle=270,width=6cm]{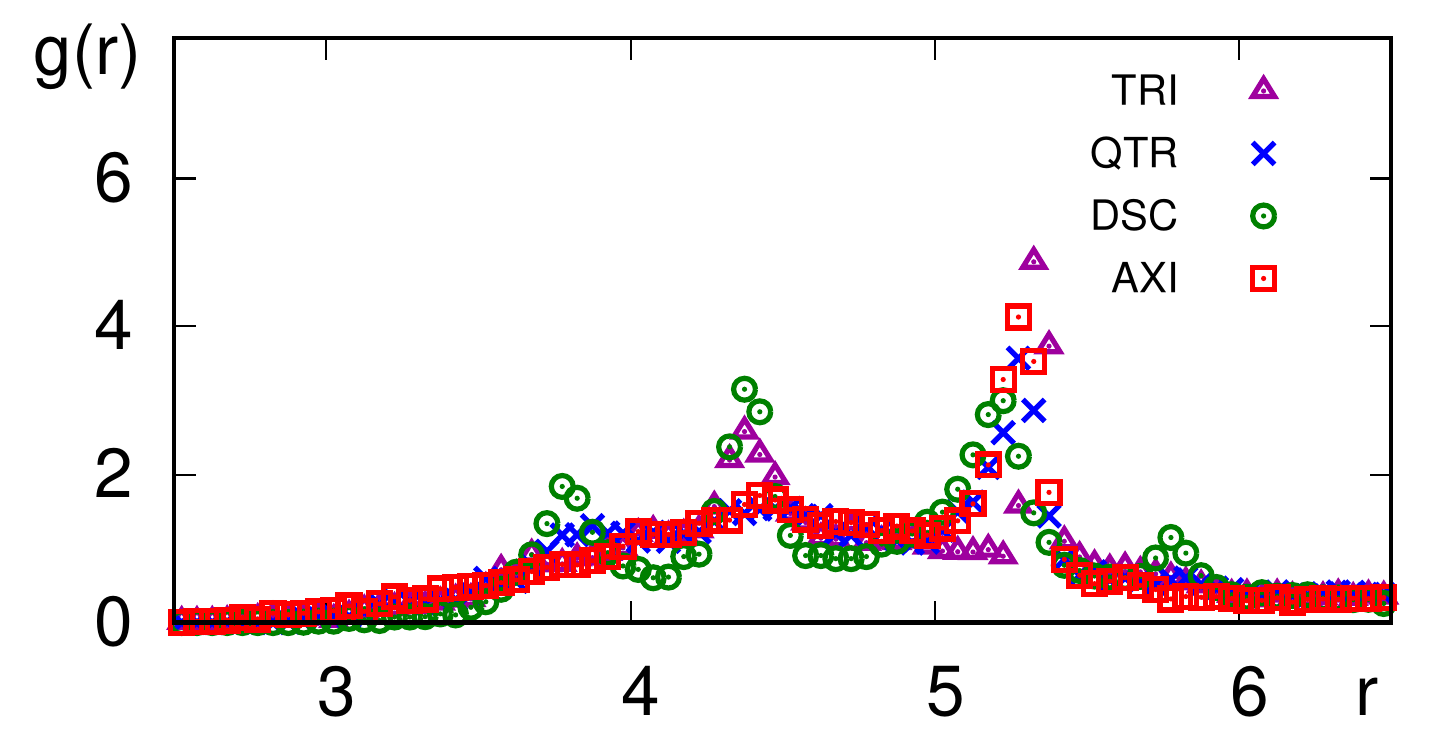}
\caption{\label{RDF}(a) GNP-GNP radial distribution functions for four patching patterns of the solution
of $N\idx{mol}=27$ PGNPs. The data in a gel state is shown via different color dotted curves indicating
respective pattern. The dispersed state data is shown as gray solid curves. (b) The same but for the
case of $N\idx{mol}=45$ PGNPs.}
\end{center}
\end{figure}
Plotting the radial distribution functions (RDFs) is a standard way to analyze the structure of liquid,
gel and solid phases. This was the case, for instance, for the self-assembled morphologies in bulk
systems of decorated GNPs.\cite{Ilnytskyi2010} We found, however, that these characteristics are much less
descriptive towards the differences in a local structure of the networks observed in the current study.
In particular, let us consider the GNP-GNP RDFs $g(r)$ for the the case of $N\idx{mol}=27$ shown in
Fig.~\ref{RDF} (a). Let us note that the Figs.~\ref{K_C_Nmol} and \ref{E_Nmol} indicate very different
values for the connectivity related characteristics $K$, $C$ and $E$ of the network for four patching
patterns TRI, QTR, DSC and AXI. One, however, observes a very minor differences in the appearance of their
respective RDFs $g(r)$ in a gel state as seen in Fig.~\ref{RDF} (a). All RDFs display a single peak
of a similar height and at a similar position of $r=5.2-5.4\nm$, which corresponds to the average
length of the ligand-ligand link. At $r<4.2$ one observes $g(r)\approx 0$, as a consequence of a strong repulsion
between PGNPs, whereas at $r>6$ $g(r)\to 1$ in the same way for all patching patterns. Upon 
an increase of the
PGNPs concentration, e.g. the case of $N\idx{mol}=45$, the interval of non zero values for RDF became much broader
and starts at $r>3\nm$, see Fig.~\ref{RDF} (b). One should note that for the QTR and AXI decoration, no peaks
are observed at $r<5\nm$ indicating relatively uniform compression of PGNPs due to the increase of their number
compared to the case of $N\idx{mol}=27$. For the TRI and EQU patching patterns, one or two peaks are observed,
at $r\approx 3.8$ and $4.35$, respectively, indicating the presence of some specific energy favorable mutual
arrangement of adjacent PGNPs.

\section{\label{V}Conclusions}

We discuss gelation in the solution of gold nanoparticles that are decorated by the ligands containing liquid crystalline
groups. The solution is confined in a slit-like pore with its walls formed of a layer of frozen particles.
Crosllinking is of soft nature, due to strong side-by-side interaction between the liquid crystalline
groups of adjacent nanoparticles. The focus of the study is on the role played by a decoration patching pattern
in the properties of gel while keeping grafting density of ligands the same in all cases. The properties of interest
include maximum subnet size $S\idx{max}$ and its wall-to-wall span $Z\idx{max}$, average vertex rank
$K$, local clustering coefficient $C$ and the wall-to-wall effective spring constant $E$ of the gel. All these
are considered in relation to the potential catalytic applications.

We found that out of six patching patterns considered, abbreviated as ROD, TRI, QTR, DSC, AXI and HDG, see
Fig.~\ref{model}, only three: TRI, DSC and AXI are capable of forming the single, wall-to-wall percolating
network gel for the conditions and system size being considered. Nevertheless, the values of $K$, $C$ and $E$
for the gels formed in the latter three cases are different in both their evolution towards gel state and their
dependence on the concentration of solution.

Mechanical stability of a gel, given by its $E$ value, is vital for its repetitive use in catalytic applications.
It depends
on both the details of interparticle connectivity and on the ` `strength'' of each link given 
by a number of pairs of
ligands it involves. As fas as the number of ligands in a single patch differs from one pattern to another, the
competition takes place between these two factors leading to a non-trivial dependence of $E$ on the concentration
of a solution which is found in this study.  We found TRI and AXI patterns to be the most suited for
forming stable gels at least in the conditions of the system under analysis.

There are many possible extensions of this study. The first one to mention is to consider dependence of gelation
process on the rigidity of the patterns on the nanoparticle surface thus mimicking more flexible attachment of
ligands. The second one is the dependence of the gelation and free area on the nanoparticles on the decoration
density, which affects possible catalytic activity. The third one is the possibility to photo-control gelation.
Finally, it would be beneficial to perform more rigorous
fine-tuning of the effective coarse-grained potentials between the building blocks of the model nanoparticles
to match the cases of some particular experimental systems. In this way one could not only compare the
properties of gels formed by differently decorated nanoparticles but also get hold on estimates of elasticity
of a gel network in real physical units.
 
\section{Acknowledgements}

J.I. acknowledges financial support by EU under FP7 IRSES Projects Nos. 612707 ``Dynamics of and in Complex Systems''
and 612669 "Structure and Evolution of Complex Systems with Applications in Physics and Life Sciences",
and S.S. -- from NCN Poland under Grant No. 2015/17/B/ST4/03615.



\balance



\end{document}